\documentclass[journal]{IEEEtran}

\usepackage{cite}      
\usepackage{siunitx}
\usepackage{array}
\usepackage{amssymb,amsmath}
\usepackage{comment}
\usepackage{hyperref}
\usepackage{tabularx}

\usepackage{graphicx}

\DeclareMathOperator*{\argmin}{arg\,min}
\DeclareMathOperator*{\argmax}{arg\,max}

\makeatother

\author{Gopalakrishnan Sundararajan,~\IEEEmembership{Student~Member,~IEEE},~Chris Winstead,~\IEEEmembership{Senior~Member,~IEEE,}~and~Emmanuel~Boutillon,~\IEEEmembership{Senior~Member,~IEEE}%
\thanks{This work was supported by the US National Science Foundation under award ECCS-0954747, and by the Franco-American Fulbright Commission for the Exchange of Scholars. The work was also used resources of the CPER PALMYRE II, with funding from FEDER and the region of Brittany, France.} %
\thanks{G.\ Sundararajan and C.\ Winstead are with the Department of Electrical and Computer Engineering, Utah State University, Logan, UT 84322-4120. Email {\em gopal.sundar@aggiemail.usu.edu} and {\em chris.winstead@usu.edu}.} %
\thanks{E.\ Boutillon is with lab-STICC, UMR 6285, Universit\'e de Bretagne Sud, 56321 Lorient, France. Email {\em emmanuel.boutillon@univ-ubs.fr}.}%
}

\newcommand{\MYheader}{\smash{\scriptsize
\hfil\parbox[t][\height][t]{\textwidth}{\centering
This is the author's version of an article that has been published in this journal. Changes were made to this version by the publisher prior to publication.\\
The final version of record is available at \url{http://dx.doi.org/10.1109/TCOMM.2014.2356458}}\hfil \thepage}}

\newcommand{\MYfooter}{\smash{\scriptsize
\hfil\parbox[t][\height][t]{\textwidth}{\centering
Copyright (c) 2014 IEEE. Personal use is permitted. For any other purposes, permission must be obtained from the IEEE by emailing pubs-permissions@ieee.org.}\hfil\hbox{}}}

\makeatletter

\def\ps@headings{%
\def\@oddhead{\MYheader}
\def\@evenhead{\MYheader}
\def\@oddfoot{\MYfooter}%
\def\@evenfoot{\MYfooter}}

\def\ps@IEEEtitlepagestyle{%
\def\@oddhead{\MYheader}%
\def\@evenhead{\MYheader}%
\def\@oddfoot{\MYfooter}%
\def\@evenfoot{\MYfooter}}

\makeatother

\begin{document}
\title{Noisy Gradient Descent Bit-Flip Decoding for LDPC Codes}

\maketitle
 
\begin{abstract}
A modified Gradient Descent Bit Flipping (GDBF) algorithm is proposed for decoding Low Density Parity Check (LDPC) codes on the binary-input additive white Gaussian noise channel. The new algorithm, called Noisy GDBF (NGDBF), introduces a random perturbation into each symbol metric at each iteration. The noise perturbation allows the algorithm to escape from undesirable local maxima, resulting in improved performance. A combination of heuristic improvements to the algorithm are proposed and evaluated. When the proposed heuristics are applied, NGDBF performs better than any previously reported GDBF variant, and comes within 0.5\, dB of the belief propagation algorithm for several tested codes. Unlike other previous GDBF algorithms that provide an escape from local maxima, the proposed algorithm uses only local, fully parallelizable operations and does not require computing a global objective function or a sort over symbol metrics, making it highly efficient in comparison. The proposed NGDBF algorithm requires channel state information which must be obtained from a signal to noise ratio (SNR) estimator. Architectural details are presented for implementing the NGDBF algorithm. Complexity analysis and  optimizations are also discussed.

\end{abstract}

\section{Introduction}
\label{Intro}

\PARstart{L}{ow} Density Parity Check (LDPC) codes gained considerable research attention in recent years. Due to their powerful decoding performance, LDPC codes are increasingly deployed in communication standards. The performance and cost of using LDPC codes are partly determined by the choice of decoding algorithm. LDPC decoding algorithms are usually iterative in nature. They operate by exchanging messages between basic processing nodes. Among the various decoding algorithms, the soft decision Belief Propagation (BP) algorithm and the approximate Min-Sum (MS) algorithm offer the best performance on the binary-input additive white Gaussian noise (AWGN) channel   \cite{MacKay_1997a,Kou_2001}, but these algorithms require a large number of arithmetic operations repeated over many iterations. These operations must be implemented with some degree of parallelism in order to support the throughput requirements of modern communication systems \cite{Mansour_2003,Guilloud_2007}. As a result, LDPC decoders can be highly complex devices. %, e.g.\ \cite{Zhang_2010a,Yen_2012,Sun_2009a,
%Mohsenin_2010a,Mansour_2006a,Jin_2008a,
%Cevrero_2010,Brandon_2009a,Blanksby_2002}.
%and incur the highest implementation cost. %Hard decision bit flipping (BF) algorithms require much less complexity at the cost of performance, with the simplest being the Gallager-A decoder. Much research has been invested into exploring the complexity/performance space that falls between the BF and SPA algorithms.

Significant effort has been invested to develop reduced-complexity decoding algorithms known ``bit-flipping'' decoders.
These algorithms are similar in complexity to hard-decision decoding algorithms, but obtain improved performance by accounting for soft channel information. %Several variants of the WBF algorithm have been reported to date.%, e.g.\ \cite{Zhang_2004,Jiang_2005}. 
%A particularly successful class of bit-flipping algorithms are the various Weighted Bit-Flipping (WBF) decoders introduced by Kou et al \cite{Kou_2001}.
In most bit-flipping algorithms, the symbol node updates are governed by an {\em inversion function} that estimates the reliability of received channel samples. The inversion function includes the received channel information in addition to the hard-decision syndrome components obtained from the code's parity-check equations. In the so-called {\em single} bit-flipping algorithms, the least reliable bit is flipped during each iteration. In {\em multiple} bit-flipping algorithms, any bit is flipped if its reliability falls below a designated threshold, hence multiple bits may be flipped in parallel, allowing for faster operation.

A recently emerged branch of the bit-flipping family is Gradient Descent Bit Flipping (GDBF), which formulates the inversion function as a gradient descent problem. 
GDBF algorithms demonstrate a favorable tradeoff between performance and complexity relative to other bit-flipping algorithms. One difficulty for GDBF algorithms is that they are affected by undesirable local maxima which cause the decoder to converge on an erroneous message. Various schemes have been proposed to avoid or escape local maxima, but require additional complexity due to multiple thresholds or computing a global function over the code's entire block length.  In this work, we propose an improved version of the GDBF algorithm, called Noisy GDBF (NGDBF) that offers a low-complexity solution to escape spurious local maxima. The proposed method works by introducing a random perturbation to the inversion function. The resulting algorithm provides improved performance and requires only local operations that can be executed fully in parallel (except for a global binary stopping condition). The proposed NGDBF algorithm comprises a set of heuristic methods that are empirically found to provide good performance for typical codes. Simulation results indicate that the NGDBF's optimal noise variance is proportional to the channel noise variance. This introduces a possible drawback compared to previous GDBF algorithms: NGDBF requires knowledge of the channel noise variance, which must be obtained from an estimator external to the decoder. Because of the heuristic nature of these results, the paper is organized to present the algorithm's technical details and empirical results first, followed by theoretical analyses that provide explanations for some of the observed results.

The remainder of this paper is organized as follows:
Section \ref{related} discusses the related work on bit-flipping algorithms, as well as some recently reported decoding algorithms that benefit from noise perturbations.
Section \ref{proposed} describes notation and summarizes the proposed NGDBF algorithm and its heuristic modifications. Section  \ref{res} presents simulation results, and offers a comparative analysis of the various heuristics. Section \ref{sec:architecture} presents architectural simplifications and complexity analysis. Section  \ref{sec:convergence_analysis} presents an evaluation and comparison of the algorithms' convergence to the global maximum-likelihood (ML) solution, and Section \ref{sec:ML_interpretation} presents an analysis of some of NGDBF's heuristics --- namely threshold adaptation and syndrome weighting --- which are interpreted in terms of evolving ML decisions on the symbol nodes' local neighborhoods. Conclusions are presented in Section \ref{conclude}.

\section{Related Work}
\label{related}

This section presents a review of bit-flipping algorithms and other methods related to the new NGDBF algorithms described in this article. As an aid to the reader, a qualitative summary of the considered bit-flipping algorithms and their comparative characteristics are provided in Table \ref{tbl:comparison}. The performance and complexity comparisons in Table \ref{tbl:comparison} are qualitative estimates made solely within the family of bit-flipping algorithms.

The original bit-flipping algorithm (BFA) was introduced by Gallager in his seminal work on LDPC codes \cite{Gallager_1963}. Gallager's BFA is a hard-decision algorithm for decoding on the binary symmetric channel (BSC), in which only hard channel bits are available to the decoder. To correct errors, the BFA computes a sum over the adjacent parity-check equations for each bit in the code. If, for any bit, the number of adjacent parity violations exceeds a specified threshold, then the bit is flipped. This process is repeated until all parity checks are satisfied, or until a maximum iteration limit is reached. The BFA has very low complexity since it only requires, in each iteration, a summation over binary parity-check values for each symbol; however the BFA provides weak decoding performance. Miladinovic et al.\ considered a probabilistic BFA (PBFA) which adds randomness to the bit-flip decision, resulting in improved performance \cite{miladinovic2005improved}. In PBFA, when a bit's parity-check sum crosses the flip threshold, it is flipped with probability $p$. The parameter $p$ is optimized empirically and is adapted toward 1 during successive iterations. 

Kou et al.\ introduced the Weighted Bit-Flipping (WBF) algorithm which improves performance over the BFA by incorporating soft channel information, making it better suited for use on   the Additive White Gaussian Noise (AWGN) channel and other soft-information channels \cite{Kou_2001}. In the WBF algorithm, all parity-check results are weighted by a magnitude that indicates reliability. For each parity-check, the weight value is obtained by finding the lowest magnitude in the set of adjacent channel samples. During each iteration, a summation $E_k$ is computed over the adjacent weighted parity-check results for each symbol position $k$. The symbol with the maximum $E_k$ (or minimum, depending on convention) is flipped. The weights are only calculated once, at the start of decoding, however the WBF algorithm requires at every iteration a summation over several weights for each symbol --- a substantial increase in complexity compared to the original BFA. In addition to the increased arithmetic complexity, WBF has two major drawbacks: first, a potentially large number of iterations are required because only one bit may be flipped in each iteration. Second, the algorithm must perform a global search to find the maximum $E_k$ out of all symbols, resulting in a large latency per iteration that increases with codeword length, thereby hindering a high-throughput implementation.

Researchers introduced several improvements to the WBF. Zhang et al.\ introduced the Modified WBF (MWBF) algorithm, which obtained improved performance with a slight increase in complexity. Jiang et al.\ described another Improved MWBF (IMWBF) algorithm which offered further improvement by using the parity-check procedure from the MS algorithm to determine the parity-check weights --- another substantial increase in complexity. Both of these methods inherit the two key drawbacks associated with single-bit flipping in the WBF algorithm.

Recently, Wu et al.\ introduced a Parallel WBF (PWBF) algorithm, which reduces the drawbacks associated with single-bit flipping in the other WBF varieties \cite{Wu_2007b}. In the PWBF algorithm, the maximum (or minimum) $E_i$ metric is found within the subset of symbols associated with each parity-check. The authors of \cite{Wu_2007b} also developed a theory relating PWBF to the BP and MS algorithms, and showed that PWBF has performance comparable to IMWBF \cite{Wu_2009}. In the PWBF algorithm, it is still necessary to find the maximum $E_i$ from a set of values, which costs delay, but the set size is significantly reduced compared to the other WBF methods, and it is independent of codeword length. In spite of these improvements, PWBF retains the complex arithmetic associated with IMWBF.

To reduce the arithmetic complexity of bit-flipping algorithms, Wadayama et al.\ devised the GDBF algorithm as a gradient-descent optimization model for the ML decoding problem \cite{Wadayama_2010_TCOMM}. Based on this model, the authors of \cite{Wadayama_2010_TCOMM} obtained single-bit and multi-bit flipping algorithms that require mainly binary operations, similar to the original BFA. The GDBF methods require summation of binary parity-check values, which is less complex than the WBF algorithms that require summation over independently weighted syndrome values. The single-bit version of the GDBF algorithm (S-GDBF) requires a global search to discover the least reliable bit at each iteration. The multi-bit GDBF algorithm (M-GDBF) uses local threshold operations instead of a global search, hence achieving a faster initial convergence. In practice, the M-GDBF algorithm did not always provide stable convergence to the final solution. To improve convergence, the authors of \cite{Wadayama_2010_TCOMM} adopted a mode-switching strategy in which M-GDBF decoding is always followed by a phase of S-GDBF decoding, leveraging high-speed in the first phase and accurate convergence in the second. 

Although the mode-switching strategy provided a significant benefit, the algorithm was still subject to spurious local maxima. Wadayama et al.\ obtained further improvements by introducing a ``hybrid'' GDBF algorithm (H-GDBF) with an escape process to evade local maxima. The H-GDBF algorithm obtains performance comparable to MS, but the escape process requires evaluating a global objective function across all symbols. When the objective function crosses a specified threshold during the S-GDBF phase, the decoder switches back to M-GDBF mode, then back to S-GDBF mode, and so on until a valid result is reached. To date, H-GDBF is the best performing GDBF variant, but requires a maximum of 300 iterations to obtain its best performance, compared to 100 for M-GDBF and S-GDBF. The major disadvantages of this algorithm are its use of multiple decoding modes, the need to optimize dual thresholds for mode switching and bit flipping, the global search operation and the global objective function used for mode switching. These global operations require an arithmetic operation to be computed over the entire code length, and would be expensive to implement for practical LDPC codes with large codeword length.
%Wadayama also proposed using a small random perturbation in the H-GDBF thresholds, similar to the novel algorithm presented in this paper. In the sequel, we present a new approach that utilizes a larger random perturbation to escape from local maxima, yielding superior performance without the need for global operations or mode switching.

Several researchers proposed alternative GDBF algorithms in order to obtain fully parallel bit-flipping and improved performance. Ismail et al.\ proposed an Adaptive Threshold GDBF (AT-GDBF) algorithm that achieves good performance without the use of mode-switching, allowing for fully-parallel operation \cite{Ismail_2013}. 
The same authors also introduced an early-stopping condition (ES-AT-GDBF) that significantly reduces the average decoding iterations at lower Signal to Noise Ratio (SNR). 
Phromsa-ard et al.\ proposed a more complex Reliability-Ratio Weighted GDBF algorithm (RRWGDBF) that uses a weighted summation over syndrome components with an adaptive threshold to obtain reduced latency \cite{Phromsa_2012}. The RRWGDBF method has the drawback of increased arithmetic complexity because it performs a summation of weighted syndrome components, similar to previous WBF algorithms. 
Haga et al.\ proposed an improved multi-bit GDBF algorithm (IGDBF) that performs very close to the H-GDBF algorithm, but requires a global sort operation to determine which bits to flip \cite{Haga_2012}. 

These GDBF algorithms can be divided into two classes: First, the low-complexity class, which includes S-GDBF and M-GDBF, AT-GDBF and RRWGDBF; in this class, mode-switching M-GDBF is the best performer. For low-complexity algorithms, the typical maximum number of iterations is reported as $T=100$. Second is the high-performance class, which includes H-GDBF and IGDBF. In the high-performance class, significant arithmetic complexity is introduced and a larger number of iterations is reported, $T=300$. H-GDBF is the best performer in this class, and in this paper we consider H-GDBF as representative of the high-performance GDBF algorithms.

In this work, we propose a new Noisy GDBF algorithm with single-bit and multi-bit versions (S-NGDBF and M-NGDBF, respectively). The M-NGDBF algorithm proposed in this work employs a single threshold and also provides an escape from the neighborhood of  spurious local maxima, but does not require the mode-switching behavior used in the original M-GDBF. The proposed algorithm also avoids using any sort or maximum-value operations.
%, and does not require complex syndrome weights like those used in RRWGDBF. 
When using the threshold adaptation procedure borrowed from AT-GDBF, as described in Section  \ref{sec:threshold_adaptation}, the proposed M-NGDBF achieves performance close to the H-GDBF and IGDBF methods at high SNR, with a similar number of iterations. We also introduce a new method called {\em Smoothed} M-NGDBF (SM-NGDBF) that contributes an additional $0.3\,{\rm dB}$ gain at the cost of additional iterations. It should be noted that Wadayama et al.\ proposed using a small random perturbation in the H-GDBF thresholds \cite{Wadayama_2010_TCOMM}; the NGDBF methods use a larger perturbation in combination with other heuristics to obtain good performance with very low complexity.

Because of its reliance on pseudo-random noise and single-bit messages, the proposed NGDBF algorithms bear some resemblance to the family of stochastic iterative decoders that were first introduced by Gaudet and Rapley \cite{gaudet2003iterative}. One of the authors (Winstead) introduced stochastic decoding for codes with loopy factor graphs \cite{Winstead_ISIT_2005}, and Sharifi-Tehrani et al.\ later demonstrated stochastic decoding for LDPC codes \cite{sharifi2006stochastic, sharifi2008fully}. High throughput stochastic decoders have been more recently demonstrated by Sharifi Tehrani et al. \cite{sharifi2010majority,sharifi2011tracking,Tehrani_TSP_2010} and by Onizawa et al. \cite{onizawa2013clockless}. Stochastic decoders are known to have performance very close to BP, allow for fully-parallel implementations, and use very simple arithmetic while exchanging single-bit messages. They may therefore serve as an appropriate benchmark for comparing complexity against the proposed SM-NGDBF algorithm (an analysis of comparative complexity is presented in Section \ref{sub:complexity_analysis}).

In addition to recent work on low-complexity decoding, there has also been some exploration of noise-perturbed decoding using traditional MS and BP algorithms. Leduc et al.\ demonstrated a beneficial effect of noise perturbations for the BP algorithm, using a method called dithered belief propagation \cite{Leduc_2012}. Kameni Ngassa et al.\ examined the effect of noise perturbations on MS decoders and found beneficial effects under certain conditions \cite{Ngassa_2014}. The authors of \cite{Ngassa_2013} offered the conjecture that noise perturbations assist the MS algorithm in escaping from spurious fixed-point attractors, similar to the hypothesis offered in this paper to motivate the NGDBF algorithm. 

Up to now, there is not yet a developed body of theory for analyzing noise-perturbed decoding algorithms, and the recent research on this topic tends to adopt a heuristic approach. In this paper we also adopt the heuristic approach, and demonstrate through empirical analysis that noise perturbations improve the performance of GDBF decoders.

\begin{table}
\renewcommand{\arraystretch}{1.3}
\centering
\caption{Summary and Comparison of Bit-Flipping Algorithms}
\label{tbl:comparison}
\scalebox{0.9}{
\begin{tabular}{llll}
\hline
Algorithm & Performance & Architecture & Arithmetic Complexity \\
\hline
\hline
BFA \cite{Gallager_1963} & Poor & Parallel & Minimum \\
PBFA \cite{miladinovic2005improved} & Fair & Parallel & Low \\
\hline
WBF \cite{Kou_2001} & Fair & Serial & Moderate \\
MWBF \cite{Zhang_2004} & Good & Serial & Moderate \\
IMWBF \cite{Jiang_2005} & Excellent & Serial & High \\
PWBF \cite{Wu_2007b} & Excellent & Parallel & High \\
\hline
S-GDBF \cite{Wadayama_2010_TCOMM} & Fair & Serial & Low \\
M-GDBF \cite{Wadayama_2010_TCOMM} & Good & Mixed & Low \\
H-GDBF \cite{Wadayama_2010_TCOMM} & Excellent & Mixed & High \\
AT-GDBF \cite{Ismail_2013} & Good & Parallel & Low \\
IGDBF \cite{Haga_2012} & Excellent & Mixed & Moderate \\
RRWGDBF \cite{Phromsa_2012} & Excellent & Parallel & High \\
\hline
Stoch.~MTFM \cite{sharifi2010majority} & Excellent & Parallel & Low \\
\hline
*S-NGDBF & Fair & Serial & Low \\
*M-NGDBF & Good & Parallel & Low \\
*SM-NGDBF & Excellent & Parallel & Low-Moderate \\
\hline
\end{tabular}
}
\vspace{0.1cm}

{\footnotesize * New algorithms described in this paper.}

\end{table}

\section{Proposed Noisy GDBF Algorithm}
\label{proposed}

\subsection{Notation}

Let $\mathbf{H}$ be a binary $m\times n$ parity check matrix, where $n > m \ge 1$. To $\mathbf{H}$ is associated a binary linear code defined by $\mathcal{C} \triangleq \left\{ c \in F_{2}^n : \mathbf{H}\mathbf{c} = 0\right\}$, where $F_{2}$ denotes the binary Galois field. The set of bipolar codewords, $\hat{\mathcal{C}}\subseteq \left\{-1,\,+1\right\}^n$, corresponding to $\mathcal{C}$ is defined by $\hat{\mathcal{C}} \triangleq  \left\{\left(1-2c_{1}\right), \left(1-2c_{2}\right), ..., \left(1-2c_{n}\right) : c \in \mathcal{C} \right\}$. 
Symbols are transmitted over a binary input AWGN channel defined by the operation $\mathbf{y} = \hat{\mathbf{c}} + \mathbf{z}$, where $\hat{\mathbf{c}} \in \hat{\mathcal{C}}$, $\mathbf{z}$ is a vector of independent and identically distributed Gaussian random variables with zero mean and variance $N_0/2$, $N_0$ is the noise spectral density, and $\mathbf{y}$ is the vector of samples obtained at the receiver. 

We define a decision vector $\mathbf{x}\in\left\{-1,+1\right\}^n$. We say that $\mathbf{x}\left(t\right)$ is the decision vector at a specific iteration $t$, where $t$ is an integer in the range $\left[0,\, T\right]$, and $T$ is the maximum number of iterations permitted by the algorithm. In iterative bit-flipping algorithms, the decision values may be flipped one or more times during decoding.  We will often omit the dependence on $t$ when there is no ambiguity. The decision vector is initialized as the sign of received samples, i.e.\ $x_k\left(t=0\right) = {\rm sign}\left(y_k\right)$ for $k=1,\,2,\,\dots,\,n$. 

The parity-check neighborhoods are defined as $\mathcal{N}\left(i\right)\triangleq\left\{ j : h_{ij} = 1 \right\}$ for $i=1,\,2,\,\dots,\,m$, where $h_{ij}$ is the $(i,\,j)^{\rm th}$ element of the parity check matrix $\mathbf{H}$. The symbol neighborhoods are defined similarly as $\mathcal{M}\left(j\right) \triangleq \left\{ i : h_{ij} = 1 \right\}$ for $j=1,\,2,\,\dots,\,n$.  The code's parity check conditions can be expressed as bipolar syndrome components $s_i\left(t\right) \triangleq \prod_{j\in \mathcal{N}(i)}x_{j}\left(t\right)$ for $i=1,\,2,\,\dots,m$. A parity check node is said to be {\em satisfied} when its corresponding syndrome component is $s_i = +1$.

\subsection{GDBF Algorithm}
\label{sec:GDBF}
The GDBF algorithm proposed in \cite{Wadayama_2010_TCOMM} was derived by considering the maximum likelihood problem as an objective function for gradient descent optimization. The standard ML decoding problem is to find the decision vector $\mathbf{x}_{\rm ML}\in\hat{\mathcal{C}}$ that has maximum correlation with the received samples $\mathbf{y}$:
\begin{equation}
	\mathbf{x}_{\rm ML} = \argmax_{\mathbf{x}\in\hat{\mathcal{C}}} \sum_{k=1}^{n} {x}_ky_k. \label{eq:ML_problem}
\end{equation}
In order to include information from the code's parity check equations, the syndrome components are introduced as a penalty term, resulting in the objective function proposed by Wadayama et al.:
\begin{equation}
	f\left(\mathbf{x}\right) = \sum_{k=1}^{n} x_ky_k + \sum_{i=1}^{m}s_i. \label{eq:objective_function}
\end{equation}
In the GDBF algorithm, a stopping criterion is used to enforce the condition $\mathbf{x}\in \hat{\mathcal{C}}$, i.e.\ any allowable solution $\mathbf{x}$ must be a valid codeword. Under this constraint, a solution that maximizes the objective function (\ref{eq:objective_function}) is  also a solution to the ML problem defined by (\ref{eq:ML_problem}). This is because for any valid codeword $\mathbf{x}$, the summation $\sum_{i=1}^{m}s_i$ is constant and equal to $m$. Since the objective functions in (\ref{eq:ML_problem}) and (\ref{eq:objective_function}) differ only by a constant term, they must have the same maxima and minima.

By taking the partial derivative with respect to a particular symbol $x_k$, the local inversion function is obtained as
\begin{equation}
	E_k = x_k\frac{\partial f\left(\mathbf{x}\right)}{\partial x_k} = x_ky_k + \sum_{i\in{\mathcal{M}}\left(k\right)} s_i.
\end{equation}
Wadayama et al.\ showed that the objective function can be increased by flipping one or more $x_k$ with the most negative $E_k$ values. The resulting iterative maximization algorithm is described as follows:

%\vspace{3 mm}
\begin{enumerate}[\setlabelwidth{Step 4}]%[leftmargin=1.25cm,label=\textbf{Step \arabic*:}]
\item[Step 1:]  Compute syndrome components $s_i = \prod_{j\in \mathcal{N}\left(i\right)}x_{j}$, for all $i \in \left\{1, 2,...., m\right\}.$ 
If $s_{i}=+1$ for all $i$, output $x$ and stop.\newline

\item[Step 2:] Compute inversion functions. For $k \in \{1,\,2,\,\dots,\,n\}$ compute 
$$E_{k} = x_{k}y_{k} + \sum_{i\in \mathcal{M}\left(k\right)} s_i.$$
%It is also convenient to introduce the symbol $S_k = \sum_{i\in M\left(k\right)} s_i$, so that the inversion function can be written $E_{k} = x_{k}y_{k} + S_k$.

\item[Step 3:] Bit-flip operations. Perform one of the following:
\begin{enumerate}%[leftmargin=0.6cm,label=(\alph*)]
\item Single-bit version: Flip the bit $x_{k}$ for $k = \argmin_{k\in \left\{1,\,2,\,\dots,\,n\right\}} E_{k}.$ \newline

\item Multi-bit version: Flip any bits for which $E_{k} < \theta$, where $\theta \in \mathbb{R}^{-}$ is the {\em inversion threshold}.  \newline
\end{enumerate}
\item[Step 4:] Repeat steps 1 to 3 till a valid codeword is detected or maximum number of iterations is reached.
\end{enumerate}

The inversion threshold is a negative real number, i.e.\ $\theta < 0$, to ensure that only bits with negative-valued $E_k$ are flipped. The optimal value of $\theta$ is found empirically, as discussed in Section \ref{sub:threshold_sensitivity}.  The single-bit GDBF algorithm (S-GDBF) incurs a penalty in parallel implementations due to the requirement of finding the minimum from among $n$ values. The multi-bit version (M-GDBF) is trivially parallelized, but does not converge well because there tend to be large changes in the objective function after each iteration. The objective function increases rapidly during initial iterations, but is not able to obtain stable convergence unless a mechanism is introduced to reduce the flipping activity during later iterations. In this paper, we consider two such mechanisms: {\em mode-switching} and {\em adaptive thresholds}.

To improve performance, the authors of \cite{Wadayama_2010_TCOMM} proposed a mode-switching modification for M-GDBF, controlled by a parameter $\mu \in \{0,\,1\}$: During a decoding iteration, if $\mu=1$ then step 3b is executed; otherwise step 3a is executed. At the start of decoding, $\mu$ is initialized to 1. After each iteration, the global objective function (\ref{eq:objective_function}) is evaluated. If, during any iteration $t$, $f\left(\mathbf{x}(t)\right) < f\left(\mathbf{x}(t-1)\right)$, then $\mu$ is changed to 0. This modification adds complexity to the algorithm, but also significantly improves performance. In the sequel (Section \ref{sec:threshold_adaptation}), it is explained that AT-GDBF eliminates the need for mode-switching by using the strictly parallel mechanism of adaptive thresholds \cite{Ismail_2013}.

\subsection{Noisy GDBF}
\label{sec:NGDBF}

In order to provide a low-complexity mechanism to escape from local maxima in the GDBF algorithm, we propose to introduce a random perturbation in the inversion function.
Based on this approach, we modify the step 2 of the GDBF algorithm as follows: 

\begin{enumerate}[\setlabelwidth{Step 2}]
\item[Step 2:] Symbol node update. For $k = 1,\,2,\,\dots,\,n$ compute
$$E_{k} = x_{k}y_{k} + w\sum_{i\in \mathcal{M}\left(k\right)} s_i  + q_{k},$$
where $w\in\mathbb{R}^+$ is a syndrome weight parameter and $q_{k}$ is a Gaussian distributed random variable with zero mean and variance $\sigma^2 = \eta^2N_{0}/2$, where $0 < \eta \leq 1$.  All $q_k$ are independent and identically distributed. 
\end{enumerate}
In this step, a syndrome weighting parameter $w$ is introduced. Syndrome weighting is motivated by the local maximum likelihood analysis presented in Section \ref{sec:ML_interpretation}. Typically $w$ and $\eta$ are close to one, and are found through numerical optimization. The optimal values for $w$ and $\eta$ are code dependent, and are found to be weakly SNR dependent in some cases.

Throughout this paper, we refer to this algorithm and its variants as {\em Noisy} GDBF (NGDBF). Both single-bit and multi-bit versions are possible, and are indicated as S-NGDBF and M-NGDBF, respectively. In this paper, mode-switching is never used in association with NGDBF; instead, threshold adaptation is employed as explained in the next subsection.

The perturbation variance proportional to $N_0/2$ was chosen based on an intuition that the algorithm's random search region should cover the same distance as the original perturbation introduced by the channel noise. The noise-scale parameter $\eta$ is introduced in order to fine-tune the optimal perturbation variance for each code. The effect of $\eta$ on performance is studied empirically in Section \ref{sub:noise_sensitivity}. For some codes, good performance is obtained when using a single SNR-independent value for $\eta$. In other cases, $\eta$ must be varied to get the best performance at different SNR values.

%This is consistent with the observation reported in \cite{Leduc_2012}, that the optimal perturbation variance increases with SNR. 
\begin{comment}
The weight parameter $w$ has been considered in previous literature, e.g.\ \cite{Phromsa_2012}, but in this paper we propose using a novel heuristic for syndrome weights, given by
\begin{equation}
\label{eq:syndrome_weight_heuristic}
	w_k = \alpha \frac{Y_{\rm max}}{d_v^{(k)}},
\end{equation}
where $Y_{\rm max}$ is the maximum (i.e.\ saturated) magnitude for channel samples, $d_v^{(k)}$ is the degree of the $k^{\rm th}$ symbol node, and $\alpha$ is a proportionality constant. In Section \ref{sec:ML_interpretation} we provide a theoretical argument to support (\ref{eq:syndrome_weight_heuristic}) based on a maximum likelihood analysis. 
\end{comment}

\subsection{Threshold adaptation}
\label{sec:threshold_adaptation}
Methods of threshold adaptation were previously investigated in order to improve the convergence of multi-bit flipping algorithms. In this paper we consider a local Adaptive Threshold GDBF (AT-GDBF) algorithm described by Ismail et al.\ \cite{Ismail_2013} in which a separate threshold $\theta_k$ is associated with each symbol node. For $k=1,\,2,\,\dots,\,n$, the threshold $\theta_k$ is adjusted in each iteration by adding these steps to the M-GDBF algorithm:

\begin{enumerate}[\setlabelwidth{Step 2}]
\item[Step 0:] Initialize $\theta_k\left(t=0\right) = \theta$ for all $k$, where $\theta \in \mathbb{R}^{-}$ is the global initial threshold parameter.
\item[Step 3b:] For all $k$, compute the inversion function $E_k$. If $E_k\left(t\right)\geq\theta_k\left(t\right)$, make the adjustment $\theta_k \left(t+1\right) = \theta_k\left(t\right)\lambda$, where $\lambda$ is a global adaptation parameter for which $0 < \lambda \leq 1$. If $E_k\left(t\right) < \theta_k\left(t\right)$, flip the sign of the corresponding decision $x_k$.
\end{enumerate}
In practice, the adaptation parameter $\lambda$ must be very close to one. The case $\lambda=1.0$ is equivalent to non-adaptive M-GDBF. According to the authors of \cite{Ismail_2013}, AT-GDBF obtains the same performance as M-GDBF with mode-switching, hence it enables fully parallel implementation with only local arithmetic operations. In the sequel we will show that threshold adaptation significantly improves performance in the M-NGDBF algorithm, at the cost of some additional complexity in the bit-flip operations. %In Section\ \ref{sec:quantization} it is shown that only a small number of distinct $\theta$ values are required when the channel information is quantized with limited precision.

\subsection{Output decision smoothing}

Convergence failures in the M-NGDBF algorithm may arise from excessive flipping among low-confidence symbols. This may occur as a consequence of the stochastic perturbation term. In this situation, the decoder may converge {\em in mean} to the correct codeword, but that does not guarantee that it will satisfy all parity checks at any specific time prior to the iteration limit $T$. More precisely, suppose the decoder is in an initially correct state, i.e.\ initially all $x_k = \hat{c}_k$. When the inversion function is computed for some $x_k$, there is a non-zero probability of erroneous flipping due to the noise contribution: 
\begin{equation}
	p_{f,\,k} = \Pr\left( x_k y_k + w\sum_{i\in \mathcal{M}\left(k\right)} s_i + q_k < \theta\right).
\end{equation}
Now suppose that $p_f$ is the least among the $p_{f,\,k}$ values among all symbols.
Then the probability $P_F$ that at least one erroneous flip occurs in an iteration is bounded by 
\begin{equation}
P_F \geq 1 - \left(1-p_f\right)^n.
\end{equation}
This probability approaches one as $n\rightarrow\infty$ for any $p_f>0$. For a sufficiently large code, it would be unlikely to satisfy all checks in a small number of iterations, even if all decisions are initially correct.

This problem may be compensated by introducing an up/down counter at the output of every ${x}_k$. The counter consists of a running sum $X_k$ for each of the $N$ output decisions. At the start of decoding, the counters are initialized at zero. During each decoding iteration, the counter is updated according to the rule 
	\begin{equation}
		\label{eq:smoothing}
		X_k\left(t+1\right) = X_k\left(t\right) + x_k\left(t\right)
	\end{equation}
 If the stopping criterion is met (i.e.\ all parity checks are satisfied) then ${x}_k$ is output directly. If the iteration limit $T$ is reached without satisfying the stopping condition, then the smoothed decision is  $\overline{x}_k = {\rm sign}\left(X_k\right)$. In practice, the summation in (\ref{eq:smoothing}) can be delayed until the very end of decoding. This saves activity in the up/down counter and hence reduces power consumption. Results in the sequel are obtained with summation only over the interval from $t=T-64$ up to $T$. When using this procedure, we refer to the algorithm as the ``smoothed'' M-GDBF method, or in shortened form as SM-NGDBF.

\section{Simulation Results}
\label{res}

\subsection{BER performance}
The proposed NGDBF algorithms were simulated on an AWGN channel with binary antipodal modulation using various regular LDPC codes selected from MacKay's online encyclopedia \cite{Mckay_2013} (all selected codes are partially irregular in parity-check degree, but are still considered regular codes). For each code, the NGDBF decoding parameters, including $\theta$, $\lambda$, $\eta$ and $w$, were optimized one at a time, holding fixed values for all but one parameter. The free parameter was adjusted using a successive approximation procedure, repeating the BER simulation in each trial, and iteratively shrinking the search domain until the best value was found. This procedure was repeated for each parameter to obtain good-performing parameters.

All NGDBF algorithms were evaluated for the rate 1/2 $(3,\,6)$ regular  LDPC code identified as PEGReg504x1008 in MacKay's encyclopedia, which is commonly used as a benchmark in previous papers on WBF and GDBF algorithms. Because our primary attention is directed at SM-NGDBF, additional simulations were performed to verify this algorithm on the rate 1/2 regular $(4,\,8)$ code identified as 4000.2000.4.244, and on the rate 0.9356 $(4,\,62)$ code identified as 4376.282.4.9598. Unless stated otherwise, all simulations use double precision floating-point arithmetic, channel samples are saturated at $Y_{\rm max}=2.5$, and the syndrome weighting is $w=0.75$.
%given by (\ref{eq:syndrome_weight_heuristic}) with $\alpha=0.9$.
%%% !!!NOTE!!!  In the simulation code it was discovered that the syndrome weight
%%% parameter was not being scaled by Ymax, so the reported parameters need to 
%%% be adjusted by dividing out Ymax. In this case the simulation recorded
%%%  alpha=2.25, but that needs to be corrected to alpha=(2.25/Ymax)=0.9

In each simulation, comparison results are provided for the BP algorithm with 250 iterations, for the MS algorithm with 5, 10 and 100 iterations. Additional appropriate comparisons are described for each result presented in this section. The MS results presented here represent the strict MS algorithm, i.e.\ they do not reflect performance for offset-MS or normalized-MS. For the M-GDBF results, the mode-switching procedure was used to obtain the best performance in all cases.

\begin{figure}[t]
	\centering
		\includegraphics[scale=1]{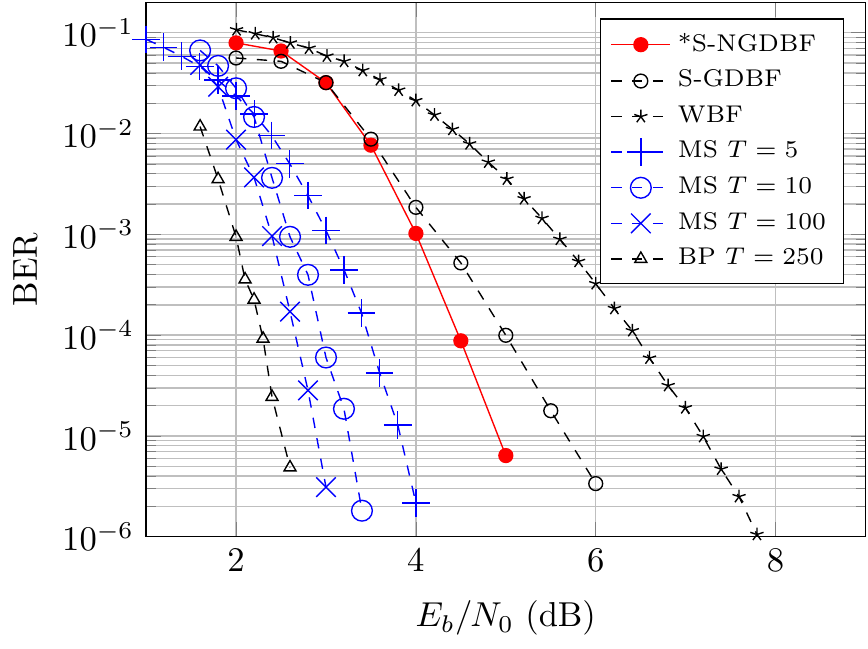}
	
	\caption{BER versus $E_{b}/N_{0}$ curves for S-NGDBF with $T=100$ and $\eta=1.0$ using the rate 1/2 PEGReg504x1008 code simulated over an AWGN channel with binary antipodal modulation. The newly proposed S-NGDBF algorithm is indicated by an asterisk (*).}
	\label{Fig:BER_SGDBF}
\end{figure}

To verify the beneficial effect of added noise, we first verified the S-NGDBF algorithm for the PEGReg504x1008 code with $T=100$. The results are shown in Fig.\ \ref{Fig:BER_SGDBF}, with comparative results for S-GDBF ($T=100$), WBF ($T=100$), MS and BP. The results show a gain approaching $1\,{\rm dB}$ for S-NGDBF compared to S-GDBF. This provides a basic empirical validation for the NGDBF concept.

\begin{figure}[tbp]
	\centering
		\includegraphics[scale=1]{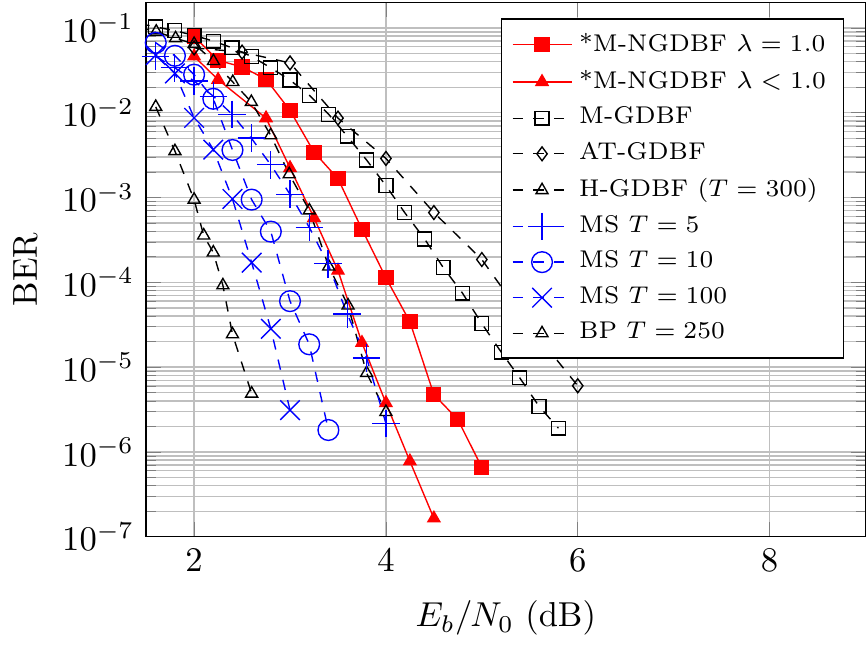}	
	\caption{BER versus $E_{b}/N_{0}$ curves for M-NGDBF with $T=100$ using the rate 1/2 PEGReg504x1008 code simulated over an AWGN channel with binary antipodal modulation. The newly proposed M-NGDBF algorithms (adaptive and non-adaptive) are indicated by asterisks (*). Several other known algorithms are shown for comparison, including M-GDBF ($T=100$), AT-GDBF ($T=100$) and H-GDBF with escape process ($T=300$).}
	
	\label{Fig:BER_MGDBF}
	
\end{figure}

Simulation results for the M-NGDBF algorithm are shown in Fig.\ \ref{Fig:BER_MGDBF}. The results in this figure were obtained for the PEGReg504x1008 code with $T=100$. The M-NGDBF results are shown for the non-adaptive case ($\lambda=1.0$) and for the adaptive-threshold case with initial threshold $\theta=-0.9$ and $\eta=0.95$, where $\eta$ is the noise-scale parameter described in Section \ref{sec:NGDBF}. For ${\rm SNR} < 3.5\,{\rm dB}$, the best performance was obtained with an adaptation parameter of $\lambda=0.99$. At higher SNR values, $\lambda$ was decreased to $0.97$ at $3.5\,{\rm dB}$, $0.94$ at $4.0\,{\rm dB}$, and $0.9$ at $4.25\,{\rm dB}$ and $4.5\,{\rm dB}$. The performance of adaptive M-NGDBF is nearly identical to that of H-GDBF with escape process, which requires $T=300$, and is also very close to MS with $T=5$.

\begin{figure}[tbp]
	\centering
		\includegraphics[scale=1]{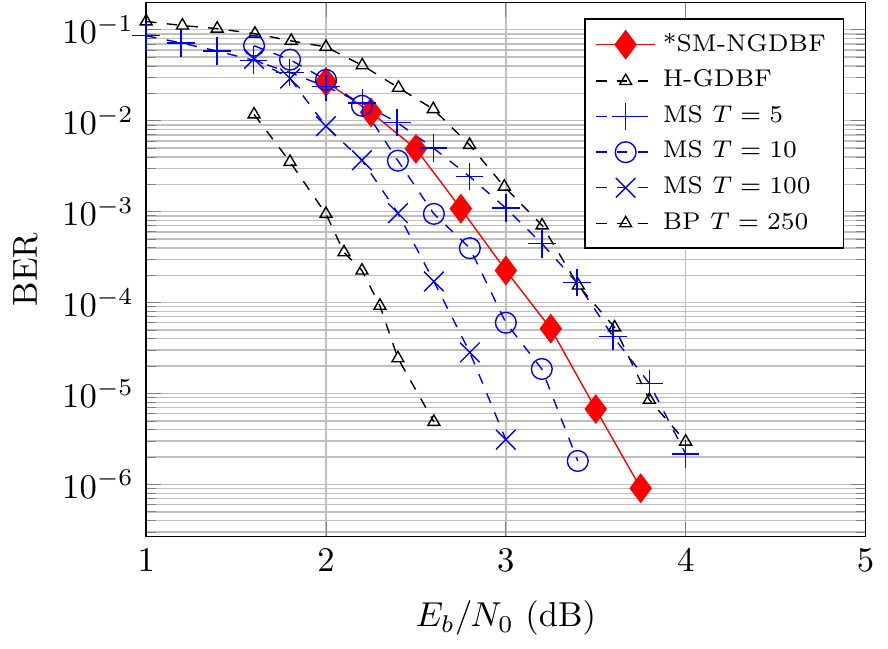}
	\caption{BER versus $E_{b}/N_{0}$ curves for SM-NGDBF and H-GDBF with $T=300$ using the rate 1/2 PEGReg504x1008 code over an AWGN channel with binary anitpodal  modulation. The proposed algorithms are indicated by an asterisk (*).}
	\label{Fig:BER_GDBF}
\end{figure}

Results for the SM-NGDBF algorithm are shown in Fig.~\ref{Fig:BER_GDBF}. For SM-NGDBF with $T=100$, performance was equal to M-NGDBF (i.e.\ there was no gain from smoothing when $T=100$), so these results are not shown. The most improved results were obtained with $T=300$, the same number of iterations used for H-GDBF with escape process. SM-NGDBF is found to achieve about $0.3\,{\rm dB}$ of coding gain compared to H-GDBF, for the same value of $T$. Compared to M-NGDBF, SM-NGDBF is found to achieve about $0.3\,{\rm dB}$ of coding gain, at the cost of additional iterations ($T=300$ for SM-NGDBF vs $T=100$ for M-NGDBF).

\begin{figure}[tbp]
	\centering
		\includegraphics[scale=1]{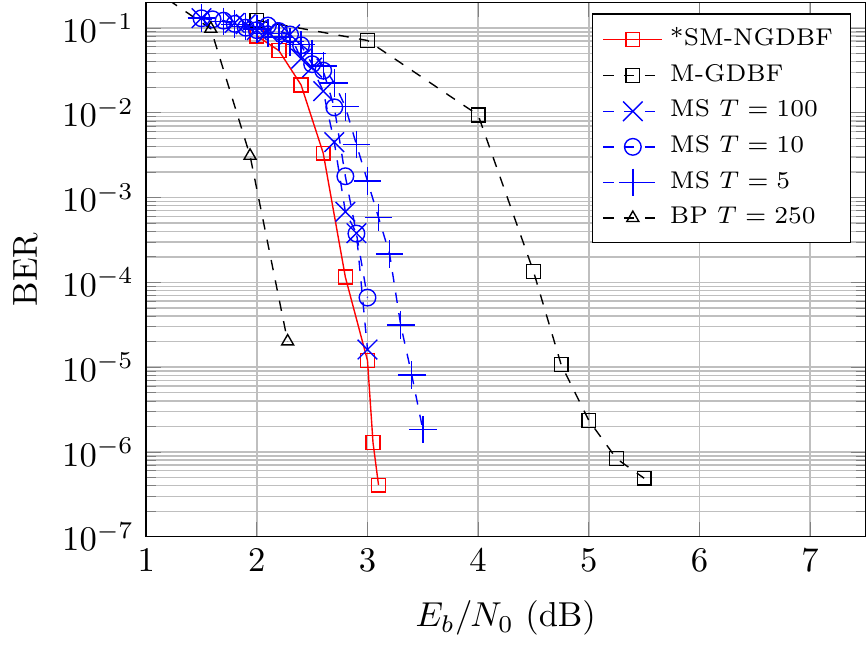}
	\caption{BER versus $E_{b}/N_{0}$ curves for SM-NGDBF with $T=300$ using the rate 1/2 4000.2000.4.244 code over an AWGN channel with binary antipodal modulation. These results were obtained using $\theta=-0.9$, $\lambda=0.99$, and $\eta$ varied between $0.625$ at low SNR (below 2.8) and $0.7$ at higher SNR (above 2.8).}
	\label{Fig:BER_GDBF4_8}
\end{figure}

\begin{figure}[tbp]
	\centering
		\includegraphics[scale=1]{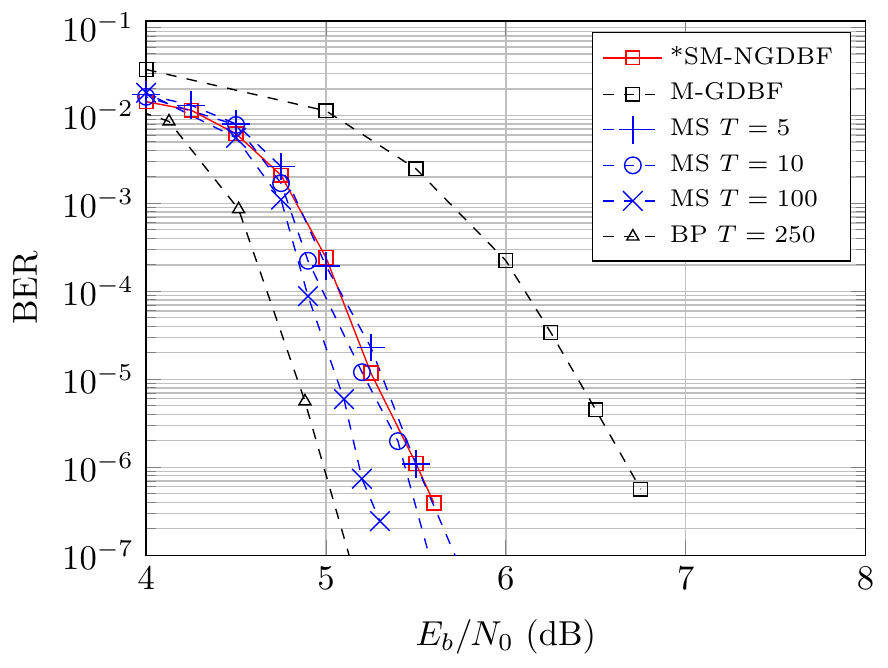}
	\caption{BER versus $E_{b}/N_{0}$ curves for SM-NGDBF with $T=300$ using the rate 0.9356 4376.282.4.9598 code over an AWGN channel with binary antipodal modulation. Several other algorithms are also shown for comparison. These results were obtained using $\theta=-0.7$, $\eta=0.65$ and $\lambda=0.993$. The dynamic range and syndrome weighting were also modified for this simulation, using $Y_{\rm max}=2.0$ and $w=0.1875$.} %$\alpha=0.375$.}
	%%% !!!NOTE!!! Here again alpha needed a correction from the original 0.75 
	%%% to 0.75/2.0=0.375 (see NOTE in previous alpha value reported above)
	\label{Fig:BER_GDBF_HighRate}
\end{figure}

In order to confirm robust performance of the SM-NGDBF algorithm, it was simulated for two other LDPC codes, yielding the results shown in Figs.~\ref{Fig:BER_GDBF4_8} and \ref{Fig:BER_GDBF_HighRate}. These results confirm that SM-NGDBF achieves good performance on codes with higher variable-node degree (in the case of Fig.~ \ref{Fig:BER_GDBF4_8}) and for codes with rates above 0.9 (in the case of Fig.~\ref{Fig:BER_GDBF_HighRate}). In both cases, SM-NGDBF remains competitive with MS decoding. 

Among the previously reported bit-flip algorithms, the best performance was achieved by Wadayama et al.'s H-GDBF algortihm with escape process. Haga and Usami's IGDBF achieved nearly identical performance to H-GDBF. Both H-GDBF and IGDBF allow a maximum of 300 iterations to achieve the best performance. Our results indicate that the adaptive M-NGDBF algorithm equals the H-GDBF performance with a maximum of only 100 iterations. Furthermore SM-NGDBF exceeds the H-GDBF performance when 300 iterations are allowed. 
These results may be interpreted in two ways. First, the adaptive M-NGDBF algorithm requires fewer iterations and is less complex than H-GDBF, but achieves the same performance (Fig.~\ref{Fig:BER_MGDBF}) --- hence it can be interpreted as a gain in speed and complexity over H-GDBF. Second, by using additional iterations with output smoothing, the speed improvement can be traded for additional coding gain (Fig.~\ref{Fig:BER_GDBF}).

\subsection{Average iterations per frame}

\begin{figure}[tbp]
	\centering	
		\includegraphics[scale=1]{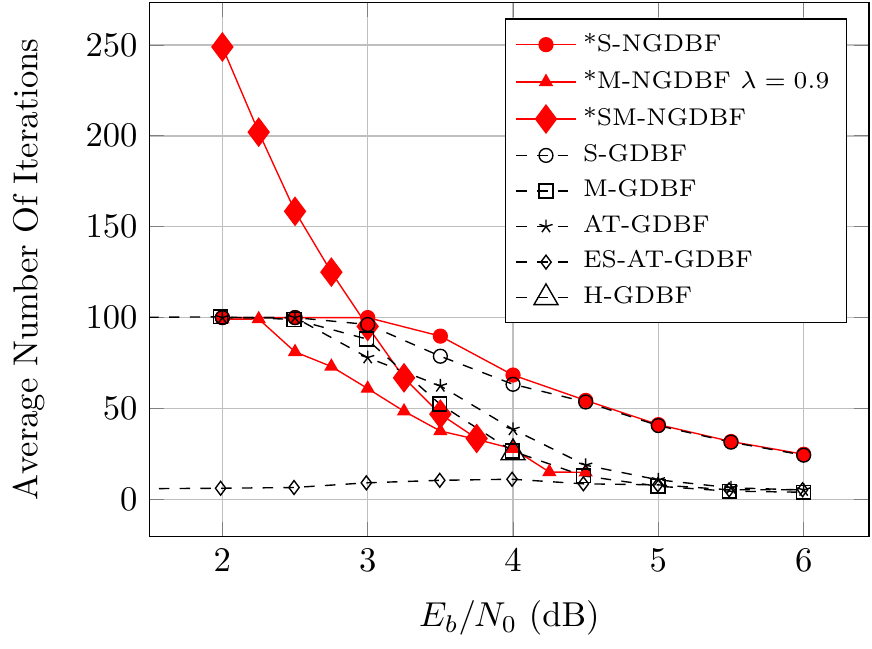}
	\caption{Average number of iterations versus $E_{b}/N_{0}$ curves for existing GDBF and  proposed NGDBF algorithms using PEGReg504x1008 code in an AWGN channel, maximum iterations limited to 100 except for SM-NGDBF and H-GDBF, which have $T=300$. The newly proposed algorithms are indicated by an asterisk (*).}
	\label{Fig:Avgite_GDBF}
\end{figure}

Fig.~\ref{Fig:Avgite_GDBF} shows the average number of iterations per frame as a function of  $E_{b}/N_{0}$, using the PEGReg504x1008 code. This plot considers results for M-NGDBF and S-GDBF with $T=100$, and for the SM-NGDBF algorithm which has $T=300$. The comparison curves show previously known GDBF algorithms with $T=100$, and also H-GDBF with $T=300$. For the H-GDBF algorithm, the full iteration profile was not disclosed, but it was stated to be 25.6 iterations at an SNR of 4$\,{\rm dB}$ \cite{Wadayama_2010_TCOMM} (shown as a single point in Fig.~\ref{Fig:Avgite_GDBF}).  From the plot, we see that the S-NGDBF provides no benefit in iteration count compared to previous algorithms. The M-NGDBF algorithms are comparable to previous  alternatives; only the Early Stopping (ES) AT-GDBF algorithm converges faster than M-NGDBF, and this advantage disappears at higher SNR. At high SNR, i.e.\ $E_b/N_0 \geq 5\,{\rm dB}$, the average iteration count is nearly the same for the M-NGDBF, SM-NGDBF, and ES-AT-GDBF methods. 

\begin{figure}[tbp]
	\centering			
		\includegraphics[scale=1]{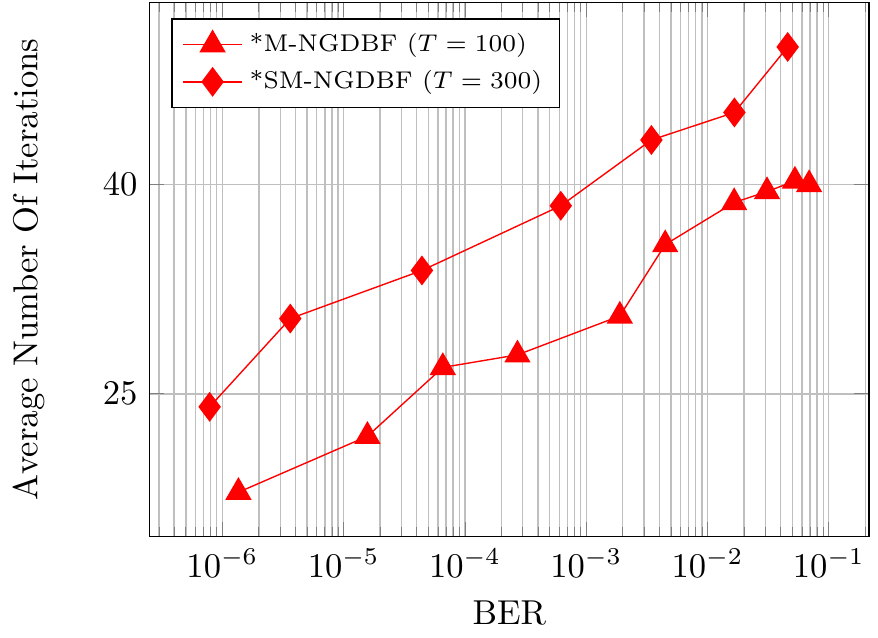}
	\caption{Average number of iterations versus BER for the proposed  M-NGDBF and SM-NGDBF algorithms.}
	\label{fig:iterations_vs_ber}
\end{figure}

At high SNR, the SM-NGDBF algorithm has the same average iterations as M-NGDBF. Although SM-NGDBF requires $T=300$ --- three times higher than M-NGDBF --- on average these algorithms require the same number of iterations when operating at the same SNR. As an alternative comparison, we compare the average number of iterations needed to achieve a given BER performance. Fig.~\ref{fig:iterations_vs_ber} shows the average iterations per frame plotted against the measured BER for M-NGDBF and SM-NGDBF. When compared for the same BER, the number of iterations needed for SM-NGDBF is on average double the number of iterations required for M-NGDBF. This result shows that the performance gain of SM-NGDBF comes at the cost of increased average iterations.

Since the average number of iterations for SM-NGDBF tends to be small, the smoothing operation is only used in a fraction of received frames. This is because the smoothing operation is only applied when the number of iterations exceeds $T-64$. For the PEGReg504x1008 code, the smoothing operation was found to be required for only 6.1\% of decoded frames when simulated at an SNR of $2.75\,{\rm dB}$, 1.45\% of frames at $3.0\,{\rm dB}$, 0.51\% of frames at $3.25\,{\rm dB}$, and 0.16\% of frames at $3.5\,{\rm dB}$.

\subsection{Sensitivity to threshold parameter $\theta$}
\label{sub:threshold_sensitivity}

\begin{figure}
	\centering
		\includegraphics[scale=1]{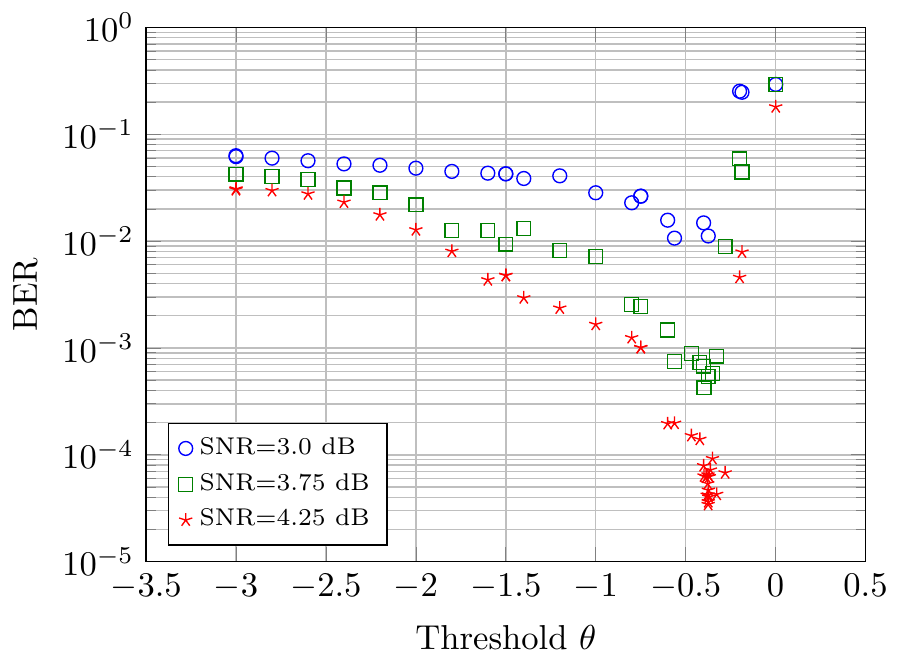}
	\caption{Threshold sensitivity of the non-adaptive M-NGDBF algorithm with parameters $\lambda=1.0$, $T=100$, $\eta=1.0$ for the PEGReg504x1008 code.}
	\label{Fig:theta_nonadaptive}
\end{figure}

The optimal threshold values for the non-adaptive M-NGDBF algorithm (i.e.\ with $\lambda=1.0$) were found empirically through a numerical search. Results from that search are shown in Fig.~\ref{Fig:theta_nonadaptive} for the PEGReg504x1008 code, in which the algorithm's BER is shown as a function of the threshold parameter. From this figure, it can be seen that the M-NGDBF algorithm is highly sensitive to the value of $\theta$, which may prove problematic if the algorithm is implemented with fixed-point arithmetic at lower precision.

\begin{figure}
	\centering
		\includegraphics[scale=1]{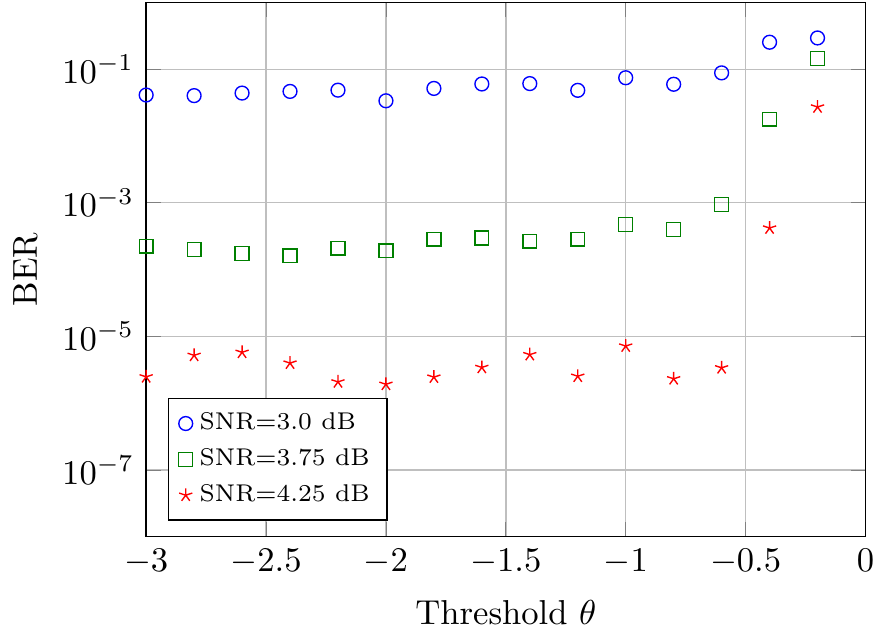}
	\caption{Threshold sensitivity of adaptive M-NGDBF algorithm with parameters $\lambda=0.9$, $T=100$, $\eta=1.0$ for the PEGReg504x1008 code.}
	\label{Fig:theta_adaptive}
\end{figure}

\begin{figure}
	\centering
		\includegraphics[scale=1]{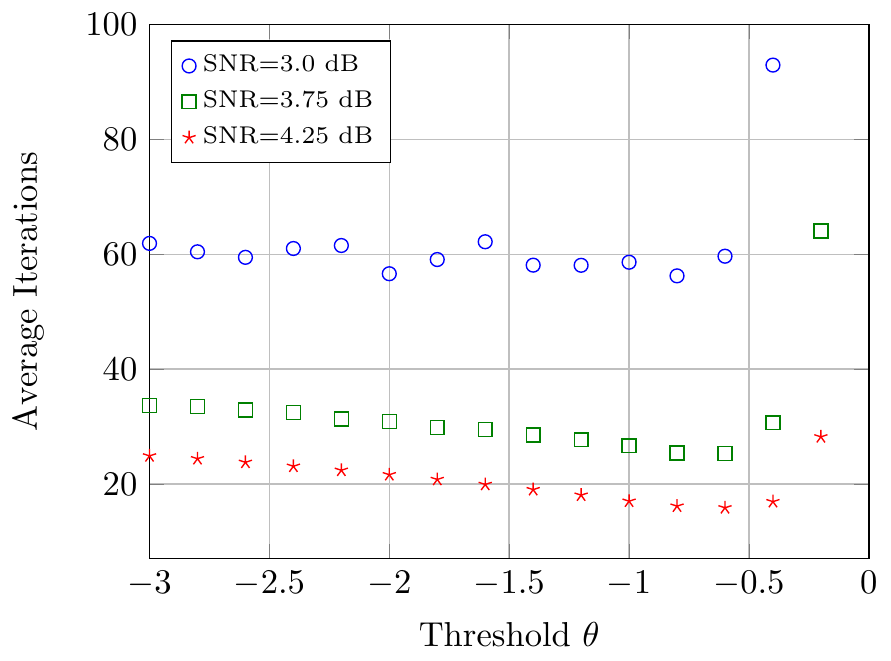}
	\caption{Average iterations for adaptive M-NGDBF as a function of the initial threshold parameter $\theta$ for the PEGReg504x1008 code. The remaining parameters are $\lambda=0.9$ and $T=100$.}
	\label{Fig:iterations_vs_theta}
\end{figure}

The adaptive M-NGDBF algorithm was simulated in a similar way, and the results shown in Fig.~\ref{Fig:theta_adaptive} reveal much less sensitivity to $\theta$. The reduced threshold sensitivity is expected because the local thresholds $\theta_k$ are iteratively adjusted during decoding. Since it will take some number of iterations for the $\theta_k$ to settle, the optimal initial threshold should be chosen as the value that minimizes the average iterations per frame. Fig.~\ref{Fig:iterations_vs_theta} shows the average number of iterations per frame as a function of $\theta$. The iteration count is seen to be only weakly a function of $\theta$, with the minimum appearing at $\theta=-0.6$.

\begin{figure}
	\centering
		\includegraphics[scale=1]{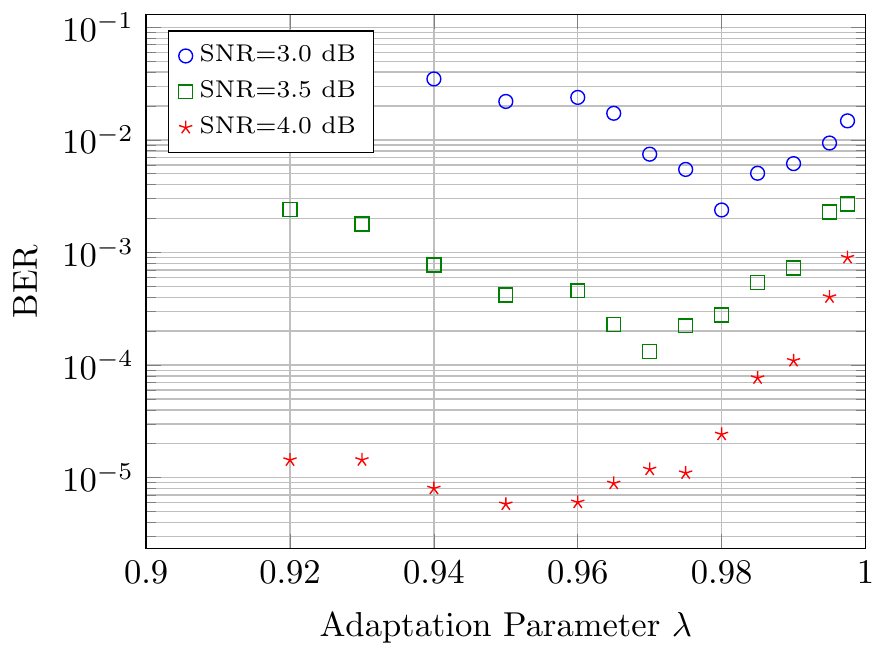}	
	\caption{Sensitivity of performance for M-NGDBF relative to the global adaptation parameter $\lambda$, with parameters $\theta=-0.9$, $T=100$, $\eta=0.96$, for the PEGReg504x1008 code.}
	\label{Fig:lambda_sensitivity}
\end{figure}

For adaptive M-NGDBF, the optimal value of $\lambda$ is found through a similar empirical search. Results from that search are shown in Fig.~\ref{Fig:lambda_sensitivity} for the PEGReg504x1008 code. These results reveal a smooth relationship between BER and $\lambda$, allowing the optimal value of $\lambda$ to be found reliably. The sensitivity revealed in Fig.~\ref{Fig:lambda_sensitivity} may prove to be difficult for implementations with quantized arithmetic; this problem is analyzed and resolved in Section \ref{sub:implementing_threshold_adaptation}.

\subsection{Sensitivity to perturbation variance}
\label{sub:noise_sensitivity}

\begin{figure}
	\centering
		\includegraphics[scale=1]{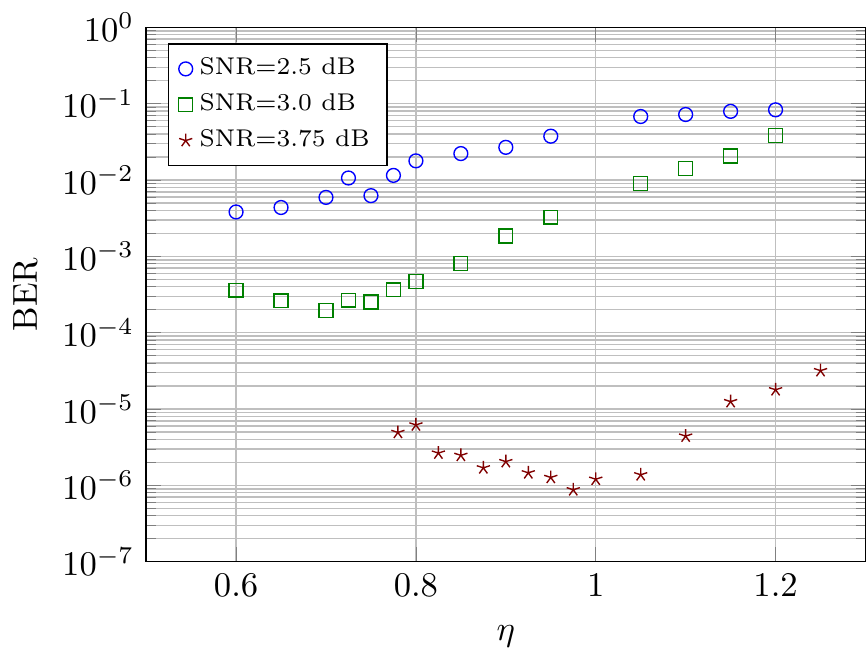}		
	\caption{Sensitivity of performance for SM-NGDBF relative to the noise scale parameter $\eta$.}
	\label{Fig:eta_sensitivity}
\end{figure}

The NGDBF algorithms' performance is sensitive to the precise variance of the noise perturbation terms. As with the $\theta$ and $\lambda$ parameters, the optimal value of the noise-scale parameter $\eta$ is found through an empirical search. This search may produce different values for different codes and at different SNR values. Example results are shown in Fig.~\ref{Fig:eta_sensitivity} for the SM-NGDBF algorithm simulated on the PEGReg504x1008 code. These results show that the optimal $\eta$ is typically somewhat less than one, and tends to increase toward one at higher SNR. 

\subsection{Effects of quantization on NGDBF}
\label{sec:quantization}
In this section we consider the performance of the M-NGDBF algorithm when implemented with limited precision. The algorithm was simulated with quantized arithmetic using $Q$ bits by applying a uniform quantization with $N_Q = 2^Q$ levels in the range $\left[-Y_{\rm max},\,Y_{\rm max}\right]$ (zero is excluded). The quantized channel sample $\tilde{y}_k$ is given by the quantization function $g\left(y\right)$:
\begin{equation}
	g\left(y\right) =  {\rm sign}\left(y\right) \left(\left\lfloor \frac{\left|y \right|  N_Q}{2Y_{\rm max}}\right\rfloor  + \frac{1}{2}\right)  \left(\frac{2Y_{\rm max}}{N_Q}\right).
	\label{eq:quantization}
\end{equation}
The quantization function is used to obtain quantized values. The vector of quantized channel samples is denoted by $\tilde{\mathbf{y}}$, and each quantized channel sample is  $\tilde{y_k} = g\left(y_k\right)$. The same function is used to obtain the quantized inversion threshold,  $\tilde{\theta} = g(\theta)$, the noise perturbation, $\tilde{q}_k = g(q_k)$, and the syndrome weight parameter, $\tilde{w} = g(w)$. After quantization, the inversion function is 
\begin{equation}
	\tilde{E}_k\left( t \right) = x_k\left( t \right) \tilde{y}_k + \tilde{w}\sum_{i\in \mathcal{M}\left(k\right)} s_i + \tilde{q}_k\left( t \right).
\end{equation}

\begin{figure}[th]
	\begin{center}
		\includegraphics[scale=1]{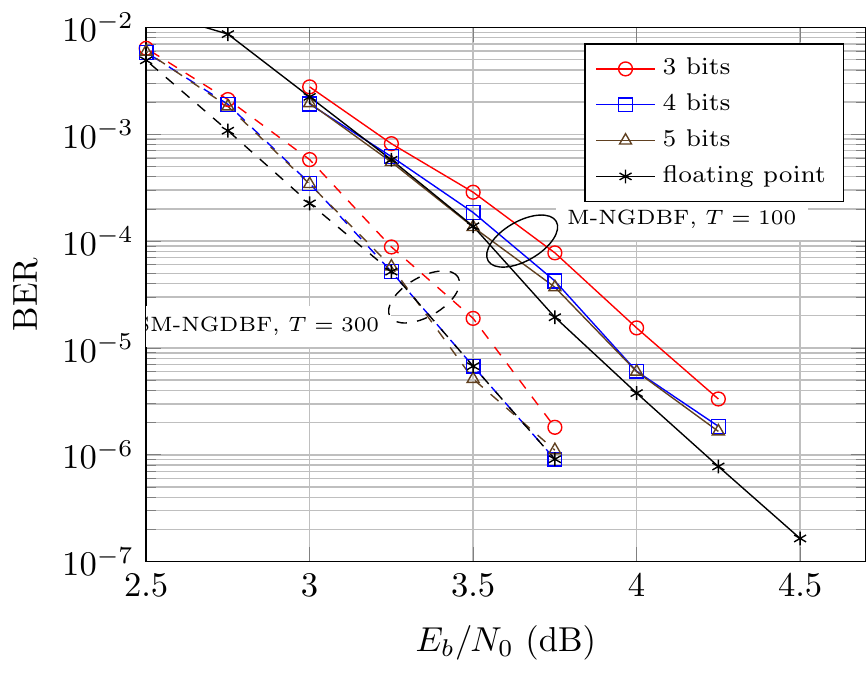}	
	\end{center}
	\caption{BER versus $E_{b}/N_{0}$ curves for quantized implementations of the proposed M-NGDBF and SM-NGDBF algorithms, using the PEGReg504x1008 code on an AWGN channel with binary antipodal modulation. Solid curves indicate results for M-GDBF with $T=100$, and dashed curves indicate results for SM-NGDBF with $T=300$.}
	\label{fig:BER_quantized}
\end{figure}

\begin{figure}[th]
	\begin{center}
		\includegraphics[scale=1]{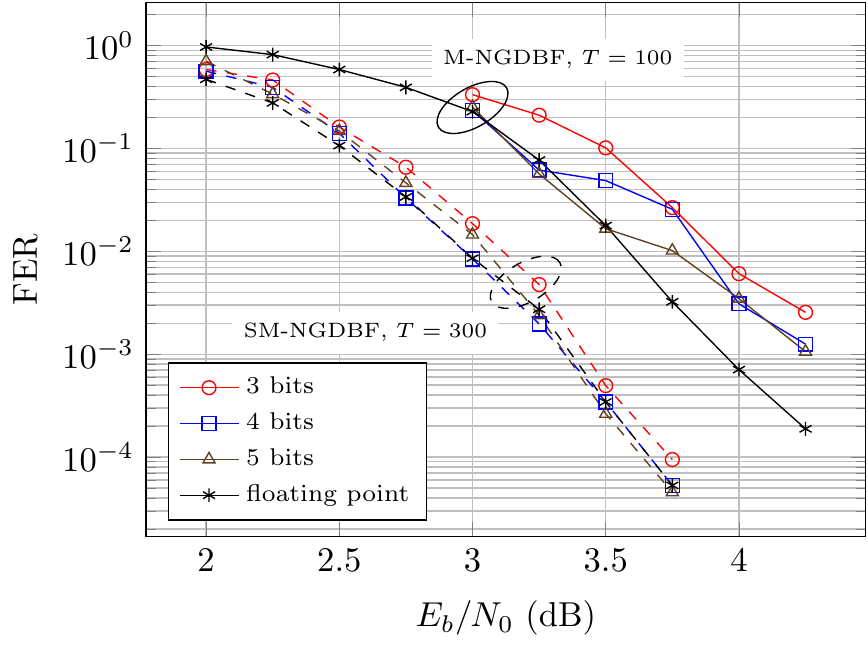}	
	\end{center}
	\caption{FER versus $E_{b}/N_{0}$ curves for quantized implementations of the proposed M-NGDBF and SM-NGDBF algorithms, using the PEGReg504x1008 code on an AWGN channel with binary antipodal modulation. Solid curves indicate results for M-GDBF with $T=100$, and dashed curves indicate results for SM-NGDBF with $T=300$.}
	\label{fig:WER_quantized}
\end{figure}

The adaptive M-NGDBF and SM-NGDBF algorithms were simulated using the quantized inversion function with the PEGReg504x1008 code. The BER results are shown in Fig.~\ref{fig:BER_quantized}. The quantized simulations reported in this section also use quantized threshold adaptation and the noise sample reuse method described in Section \ref{sec:architecture}. The results show that the algorithm is very close to unquantized performance when $Q=3$, and the best BER performance is reached when $Q=4$. There is a diminishing benefit to BER when $Q>4$, however an additional effect is observed in the Frame Error Rate (FER) results shown in Fig.~\ref{fig:WER_quantized}. Here we see an ``error flare'' effect for all cases for M-NGDBF, i.e.\ when output smoothing is not used. The flare improves when $Q$ is increased. For SM-NGDBF, i.e.\ when output smoothing is used, the flare evidently does not occur at all, or occurs at a very low FER. These simulations were performed with parameter values $Y_{\rm max}=1.7$--$1.75$, $\lambda=0.98$--$0.99$, $\theta=-0.7$ and $w=0.67$--$0.75$. %$\alpha=0.8$--$0.9$ %
%%% !!!NOTE!!! here there is another alpha correction from the original 1.4--1.5 
%%% to the corrected range 0.8--0.9.
%(the $\alpha$ parameter determines the weight parameter $w$, as explained in Section \ref{sec:ML_interpretation}). 
Parameter values were adjusted within these ranges to optimize for BER performance. 

\section{Architecture considerations}
\label{sec:architecture}

This section considers practical concerns for implementing the adaptive SM-NGDBF algorithm. These concerns include limited-precision arithmetic and architectural simplifications.

\subsection{Implementing threshold adaptation}
\label{sub:implementing_threshold_adaptation}
In Section \ref{res}, threshold adaptation was shown to provide a significant performance improvement to the M-NGDBF and SM-NGDBF algorithms. When threshold adaptation is applied, as described in Section \ref{sec:threshold_adaptation}, each symbol node must independently implement threshold scaling by parameter $\lambda$ during every iteration. If implemented with arbitrary precision, this would require implementing multiplication and division operations. When the algorithm is implemented with limited precision, however, only a small number of quantized threshold values are required. The threshold adaptation procedure can therefore be expressed as 
\begin{equation}
	\theta_k\left(t+1\right) = \left\{\begin{array}{rl}
		\theta_k\left(t\right) \lambda & x_k~\textrm{not~flipped} \\
		\theta_k\left(t\right) & x_k~\textrm{flipped} \\
	\end{array}\right.
\end{equation}
If quantized arithmetic is used with low precision, and if $\lambda$ is close to one (as is commonly the case), then it is possible that $g\left(\theta_k \lambda\right) = \theta_k$, where $g(\cdot)$ is the quantization function defined by (\ref{eq:quantization}). This case represents a failure of threshold adaptation because the local threshold is never able to change.

To avoid adaptation failures in quantized arithmetic, we introduce the symbol ${u}_k$ as a counter for non-flip events. The counter is initialized at the start of decoding as $u_k\left(t=0\right)=0$, and the counter is incremented according to the rule
\begin{equation}
	u_k\left(t+1\right) = u_k\left(t\right) + \frac{1+\delta_k\left(t\right)}{2}
\end{equation}
where $\delta_k\left(t\right) = {\rm sign}\left(\tilde{E}_k\left(t\right) - \tilde{\theta}_k\left(t\right)\right)$. %, and $l$ is substituted as a summation variable for $t$.
Then the threshold at iteration $t$ can be expressed as
\begin{equation}
	\theta_k\left(u_k\left(t\right)\right) = \theta \lambda^{{u}_k\left(t\right)},
\end{equation}
so the quantized threshold value is then given by 
\begin{equation}
	\tilde{\theta}_k\left(u_k\left(t\right)\right) = g\left(\theta\lambda^{{u}_k\left(t\right)}\right).
\end{equation}

In a finite-precision implementation, the quantized threshold $\tilde{\theta}_k\left(t\right)$ only changes for certain values of $u_k$. We say that an {\em adaptation event} occurs for some ${u}_k\left(t\right) = \tilde{\tau}$ if $\tilde{\theta}\left(\tilde{\tau}\right) \neq \tilde{\theta}\left(\tilde{\tau}-1\right)$.   Since ${u}_k$ can only be changed by zero or one during any iteration, the threshold adaptation can be implemented by storing a pre-computed list of adaptation events $\left(\tilde{\theta},\,\tilde{\tau}\right)$. 
%Whenever $\tilde{t}_k$ is equal to an adaptation level $\tilde{\tau}_j$ the corresponding $\tilde{\theta}_j$ is selected. 
Threshold adaptation can thus be implemented using a simple combinational logic circuit that detects when ${u}_k=\tau^{(i)}$ and outputs the corresponding $\tilde{\theta} = \tilde{\theta}^{(i)}$.

Table \ref{tbl:threshold_adaptation} shows threshold adaptation events for an example design with $\theta=-0.9$, $\lambda=0.99$, $T=300$ and $Y_{\rm max}=2.5$. The table shows the threshold values $\tilde{\theta}^{(i)}$ and the corresponding adaptation level $\tilde{\tau}^{(i)}$ at which the threshold value becomes active. Only two unique threshold values occur when $Q=3$; three values occur when $Q=4$; and six values occur when $Q=5$. There is typically a small number of distinct threshold values, because the values only span a small portion of the quantization range.

\def\arraystretch{1.5}
\begin{table}
\begin{center}
	\caption{Threshold adaptation events for $\theta=-0.9$, $\lambda=0.99$, $Y_{\rm max}=2.5$.}
	\label{tbl:threshold_adaptation}
	\begin{tabular}{|c|c|c|c|c|c|c|}
	\hline
	\multicolumn{3}{|c|}{$Q=3$} & \multicolumn{2}{c|}{$Q=4$} & \multicolumn{2}{c|}{$Q=5$} \\
	\hline
	$i$ & $\tilde{\theta}^{(i)}$ & $\tilde{\tau}^{(i)}$ & $\tilde{\theta}^{(i)}$ & $\tilde{\tau}^{(i)}$ & $\tilde{\theta}^{(i)}$ & $\tilde{\tau}^{(i)}$ \\
	\hline
0 & -0.9375 &0 & -0.7812 &0  & -0.8594 & 0\\
1 & -0.3125 &37 &-0.4688 &37 & -0.7031 & 15 \\
2 &  & &-0.1562 &106       & -0.5469 & 37 \\
3 &  & & & & -0.3906 & 65 \\
4 &  & & & & -0.2344 & 106 \\
5 &  & & & & -0.0781 & 175 \\
 \hline
	\end{tabular}
	\end{center}
\end{table}

\def\arraystretch{1}
\subsection{Simplification of noise sample generation}

The NGDBF algorithms require generating a Gaussian distributed random number at each symbol node during each iteration. Gaussian random number generators add significant hardware complexity. In order to simplify the implementation, we considered using only a single Gaussian Random Number Generator (RNG). The random samples are shifted from one symbol node to the next using a shift-register chain, as shown in Fig.\ \ref{Fig:architecture}. This method requires a powerup initialization so that all shift registers are pre-loaded with random samples. Simulations were performed using this method to obtain the results shown in Figs.~\ref{fig:BER_quantized} and \ref{fig:WER_quantized}, which come very close to the floating-point performance. 

As a further simplification, uniform noise samples may be used in place of Gaussian samples, but with an associated performance loss. When repeating the cases from Figs.~\ref{fig:BER_quantized} and \ref{fig:WER_quantized} using uniformly distributed noise samples, a performance loss of $0.1$--$0.2\,{\rm dB}$ was observed. Because this performance loss is undesirable, in the remainder of this paper we will only consider Gaussian distributed noise samples.

\subsection{Complexity analysis}
\label{sub:complexity_analysis}

\begin{figure}[tbh]
	\begin{center}
		\includegraphics[scale=1]{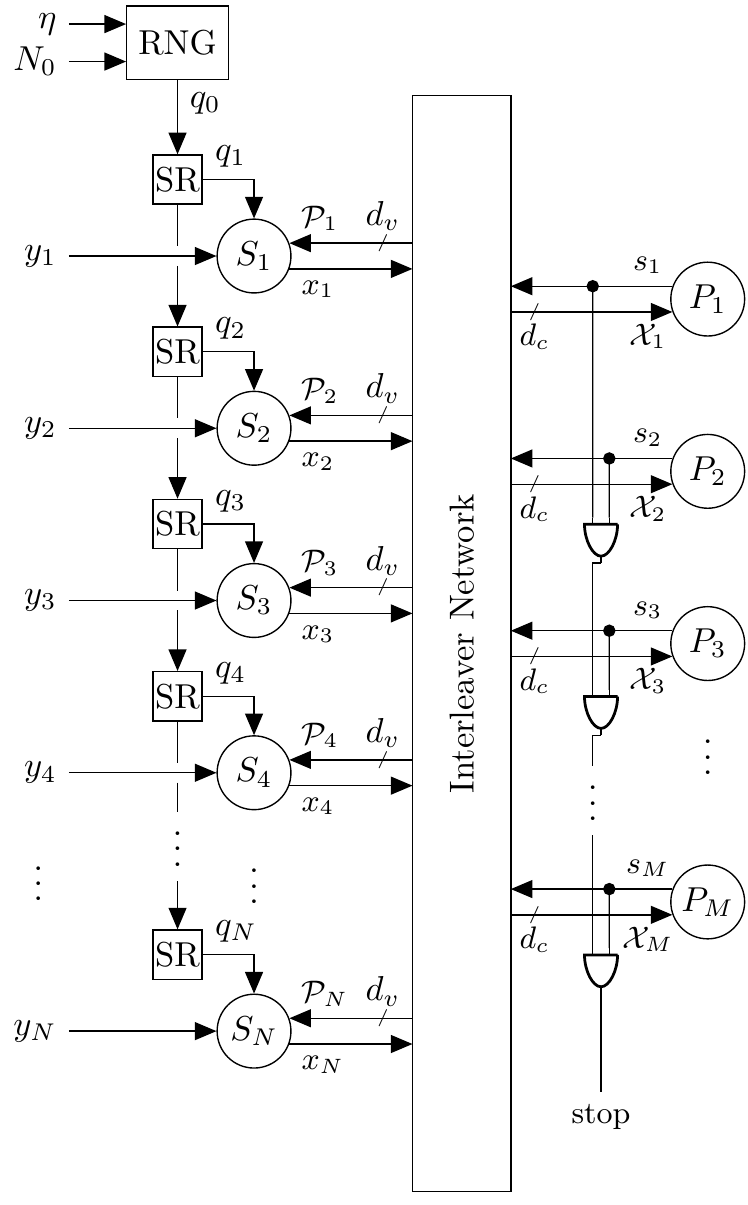}	
	\end{center}
	\caption{Architecture of the NGDBF decoder. Gaussian-distributed noise samples are produced serially at the output of a Random Number Generator (RNG). The RNG requires inputs $\eta$ and $N_0$, and the latter must be generated by a channel parameter estimator (not shown). A shift-register (SR) chain is used to distribute the random Gaussian samples that serve as the $q_k$ perturbations. The symbol $\mathcal{P}_i$ indicates the set of syndrome messages that arrive at symbol node $S_i$, corresponding to the index set $\mathcal{N}\left(i\right)$. The symbol $\mathcal{X}_j$ is the set of messages that arrive at parity-check node $P_j$, corresponding to the index set $\mathcal{M}\left(j\right)$.
	}
	\label{Fig:architecture}
\end{figure}

\begin{figure}[tbh]
	\begin{center}
			\includegraphics[scale=1]{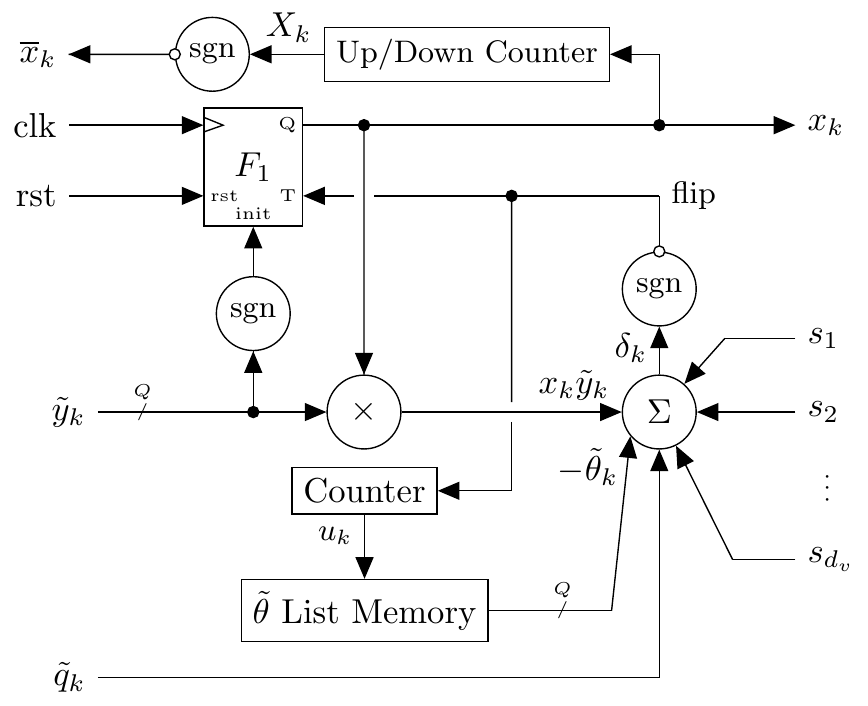}	
	\end{center}
	\caption{Symbol node schematic. $F_1$ is a toggle flip-flop. The $s_i$ messages are locally indexed. The ${\rm sgn}$ operator refers to sign-bit extraction. The multiplication $\otimes$ is binary, as it applies only to the sign bit of $\tilde{y}_k$.  }
	\label{Fig:symnode}
\end{figure}

\begin{figure}[tbh]
	\begin{center}
			\includegraphics[scale=1]{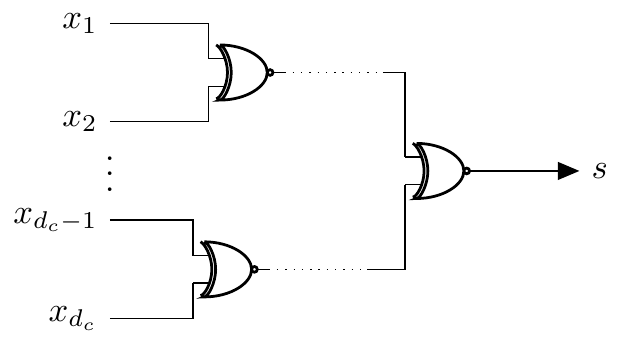}	
	\end{center}
	\caption{Check node schematic showing a tree of XNOR operations over $d_c-1$ input messages. The $x_i$ messages are locally indexed.}
	\label{Fig:chknode}
\end{figure}

The foregoing considerations are combined to arrive at the top-level architecture shown in Fig.~\ref{Fig:architecture}. In addition to the shown architecture, a channel SNR estimator is required in order to obtain $\sigma$. The symbol node implementation is shown in Fig.~\ref{Fig:symnode} and the check node implementation is shown in Fig.~\ref{Fig:chknode}. The check node implementation is uncomplicated and standard, requiring only $d_c-1$ binary XNOR operations per parity-check node. The symbol node requires one ordinary counter and one up/down counter, a $(\tilde{\theta},\,\tilde{\tau})$ memory and a signed adder with three $Q$-bit inputs and $d_v$ single-bit inputs. Four single-bit operations are also required, including a toggle flip-flop, a sign multiplier (equivalent to an XNOR operation) and two inverters. 

The most complex operation is the multi-input adder. In order to remove the weight parameter $\tilde{w}$ from the syndrome inputs, we require that all $\tilde{y}_k$, $\tilde{q}_k$ and $\tilde{\theta}$ values are pre-scaled by the factor $\tilde{w}^{-1}$ (this pre-scaling is not expressly indicated in Fig.~\ref{Fig:symnode}). Then the scaled inversion function is 
\begin{equation}
	\tilde{E}_k\tilde{w}^{-1} = x_k\tilde{y}_k\tilde{w}^{-1}  + \tilde{q}_k\tilde{w}^{-1} + \sum_{i\in {M}\left(k\right)} s_i
\end{equation}
and the flip decision can be expressed as the sign of the difference $\delta_k = \tilde{E}_k\tilde{w}^{-1} - \tilde{\theta}_k\tilde{w}^{-1}$. %; here the $\delta_k$ value is now scaled by $\tilde{w}^{-1}$ compared to its original definition in (\ref{eq:deltak}). This scaling has no effect on the algorithm since only the sign of $\delta_k$ is used in subsequent computation. 
 This detail allows for the simplified adder implementation shown in Fig.~\ref{Fig:symnode}. The effective complexity of this operation is that of two $Q$-bit binary adders and a $d_v$-bit adder.

Based on this proposed architecture, it is possible to make some high-level complexity comparisons against other related decoding methods. When making comparisons at this level, it is not possible to make strong predictions about power consumption, throughput, gate count or energy efficiency, but it is possible to make some interesting observations about the algorithms' comparative features. %The complexity comparisons are summarized in Table \ref{tbl:complexity}, corresponding to the analysis provided in the following subsections. 

\subsubsection{Comparison with previous GDBF algorithms}
Previously reported GDBF algorithms do not depend on the channel SNR, so NGDBF introduces a fixed complexity cost (i.e.\ the cost is independent of the code's length) because it requires a channel parameter estimator.
At minimum, all GDBF algorithms require $d_c-1$ XNOR operations in each parity-check node. At the symbol nodes, they require addition over the $d_v$ single-bit syndromes in each symbol node, which has gate complexity equivalent to a $d_v$-bit adder. A second $Q$-bit addition is needed to incorporate the channel information. In the S-GDBF algorithm, the minimum metric must be found, requiring $n-1$ comparisons. Since a comparison can be implemented using a signed adder, we say that the S-GDBF algorithm requires a total of $3n-1$ additions and $m\left(d_c-1\right)$ XNOR operations. In the M-GDBF algorithm, the global comparison is not required, but a comparison must still be made in each symbol node to implement the threshold operation, hence M-GDBF requires $3n$ additions. The SM-NGDBF algorithm %also requires $m\left(d_c-1\right)$ XNOR operations and 
 requires $3n$ adders, and also requires an additional $2n$ counters (a counter requires fewer gates than an adder). A single RNG module and a channel parameter estimator are also required, but these are fixed overhead that does not scale with $n$.

\subsubsection{Comparison with the MS algorithm}
The MS algorithm does not require channel SNR information. NGDBF again incurs a fixed complexity penalty due to channel parameter estimation.
In a single iteration, the MS algorithm requires at least $2d_v$ additions for every symbol node. For every parity-check node, $2d_c$ comparisons and $2d_c-1$ XNOR operations are needed. MS decoders typically allow an internal dynamic range that exceeds the channel quantization of $Q$ bits, so the arithmetic is assumed to be quantized on $Q+D$ bits, where $D\geq 0$ is the number of extra bits to accommodate the larger dynamic range. The messages exchanged between symbol and parity-check nodes are also comprised of $Q+D$ bits.

The gate requirements are clearly less for SM-NGDBF compared to MS. Based on the foregoing analysis, and using the (3,\,6) PEGReg504x1008 code as an example, SM-NGDBF requires 75\% fewer additions and comparisons. In terms of message routing, all GDBF algorithms exchange a total of $n+m$ single-bit signals in each iteration, compared to $2nd_v\left(Q+D\right)$ for MS. In the example code, assuming channel quantization with $Q=5$ and $D=3$, this means the required signal routing is reduced by 78.27\%. Based on these comparisons, we may conclude that the GDBF algorithms (including SM-NGDBF) require substantially less circuit area than MS. This analysis is not sufficient to evaluate throughput or power efficiency, since those figures depend on a variety of circuit-level considerations such as critical path delay and average switching activity in combination with the average number of iterations per frame.

Another aspect of complexity is the algorithms' decoding latency. On first inspection, we observe that the MS algorithm requires much fewer iterations than the SM-NGDBF algorithm. For example, only 4.1 iterations are needed on average for MS decoding (assuming $T=10$ with stoppping condition) on the PEGReg504x1008 code, operating at 3.5\,dB. The SM-NGDBF algorithm, at the same SNR, requires an average of 47 iterations. While it appears that latency is much greater for SM-NGDBF, we must also account for the latency per iteration in the two algorithms. In typical implementations, MS decoders utilize multiple clock cycles per iteration; for example, Zhang et al.\ used 12 clock cycles per iteration \cite{Zhang_2010a}, which we use here as a representative value. Due to SM-NGDBF's comparatively low gate and routing complexity, we expect an SM-NGDBF decoder to require only one clock cycle per iteration. We may therefore estimate the average latency of MS decoding at 49 clock cycles, compared to 47 clock cycles for SM-NGDBF. We therefore anticipate that an eventual implementation of SM-NGDBF could be comparable to previous MS implementations in terms of total latency.

\begin{comment}
, is  of the MS algorithm to SM-NGDBF and M-NGDBF algorithms is by comparing the average number of iterations needed. From Fig.~\ref{Fig:Avgite_GDBF}, we find that the average number of iterations needed for SM-NGDBF and M-NGDBF algorithms is significantly higher at low SNR values while at higher SNR values (BER $\ge$ 10E-06) the average number of iterations is less than 10. For the MS algorithm with T=10, the average number of iterations needed is same as the maximum no of iterations at all SNR values. The MS algorithm performs high latency arithmetic operations, that usually take more than a clock cycle to complete a decoding iteration \cite{Zhang_2010a}. However, the SM-NGDBF and M-NGDBF algorithms can perform a decoding operation in one clock cycle. Hence, the decoding latency in terms of number of clock cycles needed for MS algorithm  evens out in comparison to the M-NGDBF and SM-NGDBF algorithms at higher SNR values. 
\end{comment}

\subsubsection{Comparison with stochastic decoders}
The M-NGDBF algorithm bears some similarity to stochastic LDPC decoders, as was mentioned in Section \ref{related}. Stochastic LDPC decoders are known to provide performance within $0.5\,{\rm dB}$ of BP while exchanging single-bit messages with low-complexity logic processing. Stochastic decoders also require channel SNR estimation, so they share this fixed complexity cost with NGDBF. The most efficient stochastic decoding strategy is the Tracking Forecast Memory (TFM) described by Tehrani et al.\ \cite{sharifi2011tracking}. The TFM-based decoder requires $2d_c-1$ XOR operations at each parity-check node, nearly twice the number of XNOR operations needed by GDBF algorithms. At each symbol node, $2d_v$ $Q$-bit adders and $d_v$ comparisons are used, for a total of $3nd_v$ equivalent additions. Some additional supporting logic is also required, including a Linear Feedback Shift Register (LFSR) to generate random bits, and a control circuit to regulate the inputs to the TFM adder. 

Tehrani et al.\ also described a reduced-complexity Majority TFM (MTFM) design which reduces the required additions to approximately $3n$ \cite{sharifi2010majority}, making it very close to SM-NGDBF. The MTFM decoder exchanges a total of $2nd_v$ single-bit messages per iteration, compared to $n+m$ for GDBF algorithms. For the PEGReg504x1008 code, SM-NGDBF exchanges about 50\% fewer single-bit messages than an MTFM stochastic decoder for the same code. In terms of total iterations, MTFM-based stochastic decoders require about 20--40 iterations at higher SNR\cite{sharifi2010majority}, whereas SM-NGDBF requires a comparable number at 30--50 iterations for similar SNR values. We may conclude that these algorithms have very similar complexity, but SM-NGDBF should require less circuit area due to the reduced message signal routing, and because fewer XNOR operations are required.

\section{Convergence analysis}
\label{sec:convergence_analysis}
%%%%%%%%%%%%%%%%%%%%%%%%%%%%%%%%%%%%%%%

The NGDBF algorithms are built on the more general concept of noise-perturbed gradient descent optimization. 
The optimization task is the maximum likelihood (ML) decoding problem specified by (\ref{eq:ML_problem}), with the corresponding objective function $f\left(\mathbf{x}\right)$ defined by  (\ref{eq:objective_function}), as described in Section \ref{sec:GDBF}. The objective function is a non-linear function and has many local maxima. For gradient-descent optimization methods, local maxima are the  major source of sub-optimality. NGDBF rests on the hypothesis that the noisy perturbation is beneficial for escaping from local maxima, thereby improving the likelihood of obtaining the correct global maximum. This section examines that hypothesis by analyzing a detailed case example of convergence dynamics, in which NGDBF is compared to other GDBF algorithms. We expect that the algorithms' comparative convergence errors should follow the same order as their comparative BER performance.

All GDBF and NGDBF algorithms attempt to maximize $f\left(\mathbf{x}\right)$ by iteratively adjusting $\mathbf{x}$. By using the stopping condition requiring that all parity-checks are satisfied --- i.e.\ that $\prod_{i=1}^m \left(1+s_i\right)/2 = 1$ --- the GDBF algorithms enforce the constraint that candidate solutions are codewords in $\hat{\mathcal{C}}$, so long as decoding completes before reaching the maximum iteration count. For any ML-decodable case, the original transmitted codeword $\hat{\mathbf{c}}\in\hat{\mathcal{C}}$ should also be the ML solution. Then the global maximum for $f\left(\mathbf{x}\right)$ is given by
\begin{align}
\nonumber f_{\rm max} &= \displaystyle\sum\limits_{k=1}^n \hat{c}_{k}y_{k} + \displaystyle\sum\limits_{i=1}^m \prod_{j \in \mathcal{N}(i)} \hat{c}_{j} \\
&= \displaystyle\sum\limits_{k=1}^n \hat{c}_{k}y_{k} + m.
\end{align}

\begin{figure}
\centering
\includegraphics[scale=1]{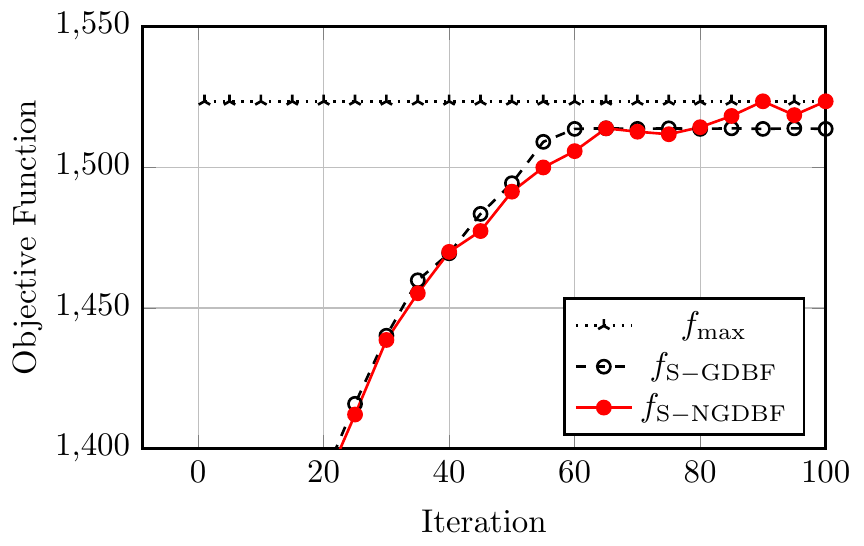}
\caption{Convergence behavior of the S-GDBF and S-NGDBF algorithms for a single frame sampled at $E_{b}/N_{0}=4\,{\rm dB}$. The true maximum is $f_{\rm max}=1523$. The S-GDBF algorithm is able to obtain a maximum value of 1514.
% algorithms are able to obtain maximum values of 1497 for AT-GDBF and 1514 for S-GDBF. 
S-NGDBF obtains the global maximum in this case.}
\label{Fig:Objective_function}
\end{figure}

\begin{figure}
\begin{center}
\includegraphics[scale=1]{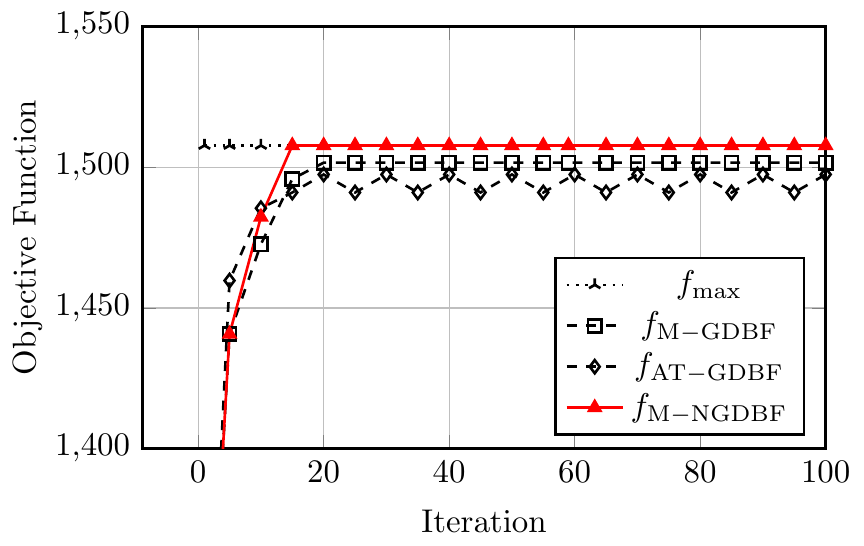}
\end{center}
\caption{Convergence behavior of the M-GDBF and M-NGDBF algorithms for a single frame sampled at $E_{b}/N_{0}=4\,{\rm dB}$. Threshold adaptation is used in the case of M-NGDBF algorithm, with $\lambda=0.99$. The true maximum is $f_{\rm max}=1508$. The M-GDBF algorithm is able to obtain a maximum value of 1502; the AT-GDBF algorithm is able to obtain a value of 1497; M-NGDBF obtains the global maximum in this case.}
\label{Fig:Objective_function2}
\end{figure}

Fig. \ref{Fig:Objective_function} shows the behavior of the objective functions evaluated for several algorithms as a function of iterations. Results are shown for the original GDBF and the proposed S-NGDBF algorithms for a simulated ML-decodable case with $E_{b}/N_{0}$ value of 4 dB. In the case of the S-GDBF algorithm, the objective function value gradually increases with the number of iterations. However, after $60$ iterations the rise eventually stops and the objective function flattens out. This flat part corresponds to a local maximum. S-NGDBF reaches the global maximum value after 90 iterations, indicating that the S-NGDBF algorithm is able to escape from the spurious local maximum.
A similar comparison is shown in Fig. \ref{Fig:Objective_function2} for the M-GDBF and M-NGDBF algorithms. The figure demonstrates that, for this example, M-GDBF is stuck in a local maximum, but M-NGDBF is able to escape from the local maximum and obtain the global solution.

The results shown in Figs.~\ref{Fig:Objective_function} and \ref{Fig:Objective_function2} represent single cases, and only partially demonstrate the superior convergence of NGDBF algorithms. To gain more insight into the convergence properties, we performed statistical analysis on the objective functions of several GDBF and NGDBF algorithms over many frames. For each algorithm, the convergence error was measured at the final iteration $T$ and averaged over $F$ transmitted frames. The convergence error is defined by
\begin{equation}
	\epsilon_{\rm alg} = \frac{1}{F}\sum_{i=1}^F\left(f_{i}\left(\mathbf{x}^{(i)}\left(T\right)\right) - f_{{\rm max},\,{i}}\right).
\end{equation}
where the subscript ``alg'' is replaced with the appropriate algorithm name, $i$ is the index of a unique sample frame, and the superscripted $\mathbf{x}^{(i)}\left(T\right)$ indicates the solution obtained for the $i^{\rm th}$ frame. The subscript $i$ is added to $f$ and $f_{\rm max}$ to emphasize their dependence on the received channel samples. For the $i^{\rm th}$ sampled frame, the transmitted message is $\hat{\mathbf{c}}^{(i)}$, the received channel samples are $\mathbf{y}^{(i)}$, and the objective function is maximized by 
\begin{equation}
	f_{{\rm max},\,i} = \sum_{k=1}^{n} \hat{c}_k^{(i)} y_k^{(i)} + m.
\end{equation}

\begin{figure}
\centering
\includegraphics[scale=1]{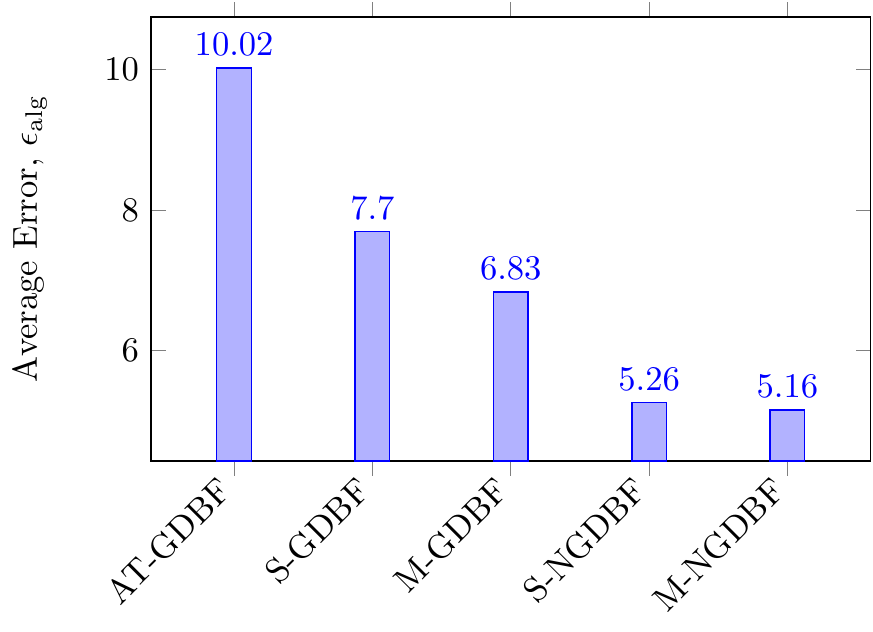}
\caption{Average convergence error for several GDBF and NGDBF algorithms at $E_{b}/N_{0} = 5\,{\rm dB}$. Threshold adaptation is used for M-NGDBF, with $\lambda=0.99$. For all algorithms, the maximum number of iterations is $T=100$.}
\label{Fig:AMSE}
\end{figure}

Fig.~\ref{Fig:AMSE} shows the average convergence error values for the tested algorithms at $E_{b}/N_{0}$ of $5\,{\rm dB}$. The total number of frames $F$ is 100. From Fig.~\ref{Fig:AMSE}, we see that the average convergence error for the proposed NGDBF algorithms is lower compared to the average error of the previously known GDBF algorithms. This shows that the NGDBF algorithms are more likely, on average, to arrive in the neighborhood of the correct solution, and are therefore more likely to have a better error correcting performance than the other GDBF algorithms.

%%%%%%%%%%%%%%%%%%%%%%%%%%%%%%%%%%%%%%%

\section{Local maximum likelihood interpretation}
\label{sec:ML_interpretation}

The GDBF and NGDBF algorithms are developed based on heuristic approaches for combinatorial optimization of the global ML objective function. In this section, we provide a theoretical analysis to motivate the use of threshold adaptation and syndrome weighting. To explain the beneficial effects of these heuristics, we consider the Local Maximum Likelihood (LML) bit-flip decision at the symbol node level given the local information from the channel and adjacent partial syndrome values. The LML analysis predicts a pattern by which the flip decisions should evolve as the decoder converges toward an error-free codeword. When using threshold adaptation and syndrome weighting heuristics with a GDBF algorithm, the evolution of flip decisions is brought into closer correspondence with the LML decisions.
During the initial iterations, LML decisions are found to be mainly determined by the channel information. In later iterations, the LML decisions are more heavily influenced by the partial syndrome values. We show that this behavior is very close to that of GDBF under threshold adaptation. We further propose that GDBF can be improved   by introducing a weight factor to the syndrome components, so that the local flip decisions evolve similarly to the LML decisions. 

In this section, we introduce one minor change in notation. Since only scalar values are considered in this section, bold-faced letters are used to indicate random variables instead of vector quantities.
We consider the problem of gradient descent decoding on a local channel sample $\tilde{y}_k$ and a set of $d_v$ adjacent syndromes $s_i$, $i\in \mathcal{M}(k)$. The channel sample is assumed to be quantized using the quantization procedure described in Section \ref{sec:quantization}. 
For a binary-input AWGN channel, we may obtain the probability masses for $\tilde{y}_k$ conditioned on the transmitted symbol $\mathbf{\hat{c}_k}$. 
\begin{align}
	\Pr\left(\tilde{y}_k \left|\,\mathbf{\hat{c}_k}=-1\right.\right) &= F_{-1}\left(\tilde{y}_k^+\right) - F_{-1}\left(\tilde{y}_k^-\right), \\
	\Pr\left(\tilde{y}_k \left|\,\mathbf{\hat{c}_k}=+1\right.\right) &= F_1\left(\tilde{y}_k^+\right) - F_1\left(\tilde{y}_k^-\right),
\end{align}
where $\tilde{y}_k^+$ and $\tilde{y}_k^-$ are the upper and lower boundary points of the quantization range that contains $\tilde{y}_k$, and $F_{-1}$ and $F_1$ are cumulative Gaussian distribution functions with variance $\sigma^2 = N_0/2$ and means $-1$ and $+1$, respectively.

Initially, the decision $x_k\left(t=0\right)$ has error probability $p_e^{\left(0\right)} = P(x_k \neq \mathbf{\hat{c}_k})$ given by $p_e=F_1\left(0\right)$.
%$p_e = \mathcal{Q}\left(-2/N_0\right)$, where $\mathcal{Q}\left(\right)$ is the tail probability of the standard normal distribution. 
%The variables associated with the local symbol node are: the transmitted symbol $\mathbf{\hat{c}}$, the current decision value $x$, and the partial syndrome messages $s_i$. 
Recalling that the syndrome values are given by $s_i = \prod_{j \in \mathcal{N}(i)} x_j$, we have $s_i = x_k\nu_{ik}$ where $\nu_{ik}$ is the partial syndrome at parity-check $i$, excluding the influence of symbol node $k$. Finally, we define $S_k$ as the penalty term $S_k=\sum_{i \in \mathcal{M}(k)} s_i$. The penalty term can also be expressed as $S_k= \left( \sum_{i \in \mathcal{M}(k)} \nu_{ik} \right)x_{k}$.

From this we directly obtain the partial syndrome error probabilities $p_c = \Pr\left(\nu_{ik} \neq \mathbf{\hat{c}_k}  \right)$ 
by enumerating over combinations in which an odd number of symbol errors has occurred out of $d_c-1$ independent, identically distributed neighbors:
\begin{equation}
	p_c  = \sum_{j=1}^{\left\lceil\frac{d_c-1}{2}\right\rceil} \binom{d_c-1}{
	2j-1} \left(1-p_e\right)^{d_c-2j}p_e^{2j-1}.
\end{equation}

The probability $P(n_e)$ of having $n_e$ errors among the partial syndrome values $\nu_{ik}$ is thus:
\begin{equation}
 P(n_e) = \binom{d_v}{n_e}p_c^{n_e}(1-p_c)^{d_v-n_e}.
\end{equation}
If $x_k = \mathbf{\hat{c}_k}$, then $n_e$ errors give a penalty $S_k = d_v - 2n_e$ (summation of $d_v-n_e$ syndrome $+1$ and $n_e$ syndromes $-1$). Symmetrically, if $x_k = -\mathbf{\hat{c}_k}$, then $S_k = 2n_e  - d_v$. In other words, knowing the observation $S_k$, we can deduce:

\begin{equation}
		P(S_k\,\left|\,x_k = \mathbf{\hat{c}_k}\right.) = P(n_e = (d_v - S_k)/2)
\end{equation}

and

\begin{equation}
		P(S_k \, \left|\, x_k = -\mathbf{\hat{c}_k}\right.) = P(n_e = (d_v + S_k)/2)
\end{equation}
Then the LML decision is 
\begin{equation}
	\hat{x}_{k,\rm LML} = \argmax_{x_k} {\rm Pr}\left(\tilde{y}_k \left|x_k\right.\right){\rm Pr}\left(S_k \left|x_k\right.\right).
\end{equation}
To relate the LML result to bit flipping algorithms, it can be expressed as an LML flip decision $\phi$, defined by
\begin{equation}
	\phi\left(x_k,\,\tilde{y_k},\,S_k\right) = {\rm sign} \log \left( \frac{\Pr\left(\tilde{y}_k \left|x_k\right.\right)\Pr\left(S_k \left|x_k\right.\right) }{\Pr\left(\tilde{y}_k \left|-x_k\right.\right)\Pr\left(S_k \left|-x_k\right.\right)} \right).
\end{equation}
If $\phi=-1$, then the optimal decision is to flip $x_k$.

In order to visualize the LML behavior on a quantized channel, we arrange the decisions in a {\em flip matrix} $\Phi$ that expresses all possible states; the rows of $\Phi$ correspond to the possible channel sample values, and the columns correspond to the possible values of $S_k$. There are $d_v+1$ possible values of $S_k$: $-d_v, -d_v+2,\dots,d_v-2,d_v$. We index these values in ascending order as $S_k^{(j)}$, $j=1,\,2,\,\dots, d_v+1$. The possible $\tilde{y}_k$ values are similarly indexed in ascending order as $\tilde{y}_k^{(i)}$, $i=1,\,2,\,\dots,N_Q$. Then $\Phi$ is an $N_Q \times (d_v+1)$ matrix with entries $\phi_{i,\,j} = \phi\left(x_k,\tilde{y}_k^{(i)},\,S_k^{(j)}\right)$.
For a given locally received $\tilde{y}_k$ and $S_k$, if the corresponding $\phi_{ij}=-1$, then the corresponding decision $x_k$ should be flipped.

\begin{figure}
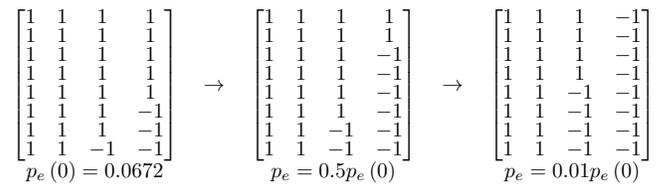

\centering
\scalebox{0.8}{
\renewcommand*{\arraystretch}{0.7}
\begin{tabular}{ccccc}
	$\begin{bmatrix}
	1 &1 &1 &1 \\
1 &1 &1 &1 \\
1 &1 &1 &1 \\
1 &1 &1 &1 \\
1 &1 &1 &1 \\
1 &1 &1 &-1 \\
1 &1 &1 &-1 \\
1 &1 &-1 &-1 \\
	\end{bmatrix}$
	
	& $\rightarrow$ &
	
	$\begin{bmatrix}
	1 &1 &1 &1 \\
1 &1 &1 &1 \\
1 &1 &1 &-1 \\
1 &1 &1 &-1 \\
1 &1 &1 &-1 \\
1 &1 &1 &-1 \\
1 &1 &-1 &-1 \\
1 &1 &-1 &-1 \\
	\end{bmatrix}$
	
	& $\rightarrow$ &
	$\begin{bmatrix}
	1 &1 &1 &-1 \\
1 &1 &1 &-1 \\
1 &1 &1 &-1 \\
1 &1 &1 &-1 \\
1 &1 &-1 &-1 \\
1 &1 &-1 &-1 \\
1 &1 &-1 &-1 \\
1 &1 &-1 &-1 \\
	\end{bmatrix}$
	\\
	
	$p_e\left(0\right)=0.0672$ & & $p_e = 0.5p_e\left(0\right)$ & & $p_e = 0.01p_e\left(0\right)$
	
\end{tabular}
}

\caption{Example of evolution of the LML flip matrix $\Phi$ for a (3,6) LDPC code, with $Q = 4$ and an $E_b/N_0 = 3.50\,{\rm dB}$. The initial value of $P_e$ is 0.0672. Since $\Phi$ is symmetric with $\Phi\left(i,\,j\right) = \Phi\left(N_Q+1-i,\,d_v+2-j\right)$, only the top $N_Q/2$ rows are shown.}
\label{fig:flip_matrix_evolution}

\end{figure}

The LML flip decision depends on the partial syndrome error probabilities, which change in successive iterations. In order to understand how the LML decision evolves across iterations, we suppose that the bit error probability is a function of the iteration number, $t$, and that $p_e\left(t\right)$ is decreasing with successive iterations. 
As the error probability decreases, the flip matrix is found to evolve from an initial pattern in which decisions are heavily dependent on $\tilde{y}_k$, with increasing dependence on $S_k$ in later iterations as $p_e{\left(t\right)}$ decreases toward zero. An example of this evolution is shown in Fig.~\ref{fig:flip_matrix_evolution}, for the case $x_k=1$ with parameters $Y_{\rm max}=1.5$, $\sigma=0.668$, $Q=4$ and $d_v=3$. For a (3,\,6) LDPC code, this corresponds to $E_b/N_0 = 3.50\,{\rm dB}$.

\begin{figure}
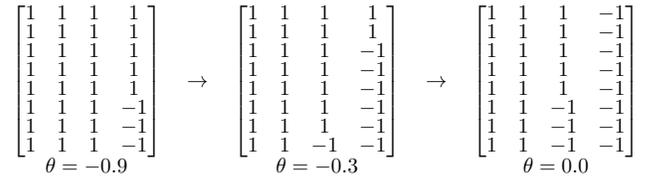

\centering
\scalebox{0.8}{
\renewcommand*{\arraystretch}{0.7}
\begin{tabular}{ccccc}
	$\begin{bmatrix}
	1 &1 &1 &1 \\
1 &1 &1 &1 \\
1 &1 &1 &1 \\
1 &1 &1 &1 \\
1 &1 &1 &1 \\
1 &1 &1 &-1 \\
1 &1 &1 &-1 \\
1 &1 &1 &-1 \\
%-1 &1 &1 &1 \\
%-1 &1 &1 &1 \\
%-1 &1 &1 &1 \\
%1 &1 &1 &1 \\
%1 &1 &1 &1 \\
%1 &1 &1 &1 \\
%1 &1 &1 &1 \\
%1 &1 &1 &1
	\end{bmatrix}$
	
	& $\rightarrow$ &
	
	$\begin{bmatrix}
	1 &1 &1 &1 \\
1 &1 &1 &1 \\
1 &1 &1 &-1 \\
1 &1 &1 &-1 \\
1 &1 &1 &-1 \\
1 &1 &1 &-1 \\
1 &1 &1 &-1 \\
1 &1 &-1 &-1 \\
%-1 &-1 &1 &1 \\
%-1 &1 &1 &1 \\
%-1 &1 &1 &1 \\
%-1 &1 &1 &1 \\
%-1 &1 &1 &1 \\
%-1 &1 &1 &1 \\
%1 &1 &1 &1 \\
%1 &1 &1 &1 
	\end{bmatrix}$
	
	& $\rightarrow$ &
	$\begin{bmatrix}
	1 &1 &1 &-1 \\
1 &1 &1 &-1 \\
1 &1 &1 &-1 \\
1 &1 &1 &-1 \\
1 &1 &1 &-1 \\
1 &1 &-1 &-1 \\
1 &1 &-1 &-1 \\
1 &1 &-1 &-1 \\
%-1 &-1 &1 &1 \\
%-1 &-1 &1 &1 \\
%-1 &-1 &1 &1 \\
%-1 &1 &1 &1 \\
%-1 &1 &1 &1 \\
%-1 &1 &1 &1 \\
%-1 &1 &1 &1 \\
%-1 &1 &1 &1
	\end{bmatrix}$
	\\
	
	$\theta = -0.9$ & & $\theta=-0.3$ & & $\theta=0.0$
	
\end{tabular}
}

\caption{Evolution of the flip matrix for the weighted GDBF algorithm with threshold adaptation. Use of threshold adaptation and syndrome weighting achieves qualitative agreement with the LML flip decisions in Fig.~\ref{fig:flip_matrix_evolution}. }

\label{fig:wGDBF_evolution}

\end{figure}

With threshold adaptation, the GDBF algorithm's behavior is similar to the LML flip matrix. Initially, the threshold $\theta$ is set to a significantly negative value, say $\theta=-1.0$. %from Section \ref{sec:ML_evolution}, 
For a given set of parameters, 
we may obtain a matrix $E$ of values for the inversion function, with members $E_{i,\,j} = \tilde{y}_k^{(i)} + S_k^{(j)}$. 
%For the same parameters used in the example shown in Fig.~\ref{fig:flip_matrix_evolution},
\begin{comment}
\begin{equation}
\tilde{E} =  \begin{bmatrix}
4.41 &2.41 &0.41 &-1.59 \\
4.22 &2.22 &0.22 &-1.78 \\
4.03 &2.03 &0.03 &-1.97 \\
3.84 &1.84 &-0.16 &-2.16 \\
3.66 &1.66 &-0.34 &-2.34 \\
3.47 &1.47 &-0.53 &-2.53 \\
3.28 &1.28 &-0.72 &-2.72 \\
3.09 &1.09 &-0.91 &-2.91 \\
-2.91 &-0.91 &1.09 &3.09 \\
-2.72 &-0.72 &1.28 &3.28 \\
-2.53 &-0.53 &1.47 &3.47 \\
-2.34 &-0.34 &1.66 &3.66 \\
-2.16 &-0.16 &1.84 &3.84 \\
-1.97 &0.03 &2.03 &4.03 \\
-1.78 &0.22 &2.22 &4.22 \\
-1.59 &0.41 &2.41 &4.41
\end{bmatrix}
\end{equation}
\end{comment}
%
By applying the threshold $\theta$ to all the elements of $E$, we obtain the flip matrix for the GDBF algorithm. As the threshold is adapted toward zero, the flip matrix evolves to place increased weight on the syndrome information, similar to the LML flip matrix evolution. The GDBF flip matrices do not correspond perfectly to the LML predictions. To bring closer agreement, a weight factor is introduced, giving a modified weighted inversion function 
\begin{equation}%
	%\tilde{E}_{k} = {x}_k\tilde{y}_k + \alpha\frac{Y_{\rm max}}{d_v}\sum_{i\in M\left(k\right)} s_i + \tilde{q}_k,
	\tilde{E}_{k} = {x}_k\tilde{y}_k + w\sum_{i\in M\left(k\right)} s_i + \tilde{q}_k,
	\label{eq:weighted_GDBF}
\end{equation}
%where $\alpha$ is a scale parameter. With this weighting, the syndrome contribution is always proportional to the maximum contribution from $\tilde{y}_k$. 
The best value for $w$ is found empirically and may be code dependent. Based on the parameters $\sigma=0.668$, $Y_{\rm max}=1.5$, $w=0.75$
%$\alpha=1.5$ 
and $d_v=3$, the flip matrix evolution corresponding to $\tilde{E}_k$ is shown in Fig.\ \ref{fig:wGDBF_evolution}. 

The relationship between LML and the GDBF heuristics is not an exact correspondence. The LML analysis predicts the desirable behavior of the flip matrix over time during a successful decoding event. When GDBF is augmented by introducing the threshold adapation and syndrome weighting heuristics, its behavior is brought into approximate correspondence with the LML prediction. This provides a new theoretical motivation for using these heuristic methods, which has not been addressed in the previous literature.

\section{Conclusions}
\label{conclude}

This paper introduced a collection of novel Noisy GDBF algorithms, based on a noisy gradient descent heuristic, that outperforms existing GBDF algorithms. We found that previous GDBF algorithms, including the S-GDBF, M-GDBF and AT-GDBF algorithms, are significantly improved when combined with the noise perturbation. Additional heuristic improvements were introduced that achieved a significant performance benefit in comparison to the best known versions of GDBF, achieving performance comparable to the standard min-sum algorithm for several LDPC codes. We also provided an architecture for implementing the new algorithm with quantized channel information, and showed that its implementation complexity is quite low compared to min-sum or stochastic decoding. The NGDBF algorithms do require estimation of the channel SNR, which introduces a fixed complexity cost that does not affect the previously known GDBF of MS algorithms.

The NGDBF decoding algorithms are based on a heuristic approach. In order to gain additional validation for those heuristics, we examined the convergence characteristics and found that, on average, the NGDBF algorithm converges closer to the global maximum whereas other algorithms are more frequently trapped in suboptimal local maxima. We also examined the approximate local ML solution for bit-flip behavior. From this analysis, we proposed using a weight factor to bring GDBF closer to the LML behavior.
As a result of these analyses, we obtained a new bit-flipping decoding algorithm that avoids using any global search or sort operations. The resulting algorithm is feasibly a competitor to the popular min-sum algorithm, since it requires less computational effort while maintaining good BER and FER performance.

\bibliographystyle{IEEEtran}
\bibliography{IEEEabrv,TCOM_preprint.bib}

\end{document}